\documentclass[useAMS]{mn2e}
\usepackage{times}
\usepackage{amssymb}
\usepackage{amsmath}
\usepackage{rotating}
\usepackage{epsfig}
\input{epsf}

\title[What Powers the Relativistic Jets in BL Lacs?]
{What Powers the Most Relativistic Jets? I: BL Lacs}

\author[E. Gardner and C. Done]
{Emma Gardner and Chris Done\\
Department of Physics, University of Durham, South Road,
Durham DH1 3LE, UK\\}

\date{Submitted to MNRAS}
\pagerange{\pageref{firstpage}--\pageref{lastpage}} \pubyear{}

\begin{document}

\topmargin = -0.5cm

\maketitle

\label{firstpage}

\begin{abstract}

The dramatic relativistic jets pointing directly at us in BL Lac
objects can be well modelled by bulk motion beaming of synchrotron
self-Compton emission powered by a low Eddington fraction accretion flow.
Nearly 500 of these AGN are seen in the 2nd
Fermi Large Area Telescope catalogue of AGN. We combine the jet models
which describe individual spectra with the expected jet parameter
scalings with mass and mass accretion rate to predict the expected
number of Fermi detected sources given the number densities of AGN
from cosmological simulations. We select only sources with Eddington
scaled mass accretion rate $< 0.01$ (i.e. radiatively inefficient
flows), and include cooling, orientation effects, and the effects of
absorption from pair production on the extragalactic IR background.

These models overpredict the number of Fermi detected BL Lacs by a
factor of 1000! This clearly shows that one of the underlying
assumptions is incorrect, almost certainly that jets do not scale
simply with mass and accretion rate. The most plausible additional
parameter which can affect the region producing the Fermi emission is
black hole spin. We can reproduce the observed numbers of BL Lacs if
such relativistic jets are only produced by the highest spin
($a_*>0.8$) black holes, in agreement with the longstanding spin-jet
paradigm.  This also requires that high spins are intrinsically rare,
as predicted by the cosmological simulations for growing black hole
mass via chaotic (randomly aligned) accretion episodes, where only the
most massive black holes have high spin due to black hole-black hole
mergers.

\end{abstract}


\section{Introduction} \label{sec:introduction}

Relativistic jets are the most dramatic consequence of accretion onto
stellar mass (black hole binaries: BHBs) and supermassive black holes
(SMBHs). Blazars are extreme examples of this, where the jet is viewed
very close to the line of sight so its emission is maximally boosted
by the relativistic bulk motion and can dominate the spectrum of the
AGN from the lowest radio energies up to TeV. However, despite years
of study, the fundamental issues of powering and launching the jets
are not understood. There is general agreement only that it requires
magnetic fields, but whether these can be generated solely from the
accretion flow or whether the jets also
tap the spin energy of the black hole (Blandford \& Znajek 1977) is
still an open question. It is also difficult to test this
observationally as neither black hole spin nor total jet power are 
easy to measure, leading to divergent views e.g. in BHBs compare
Russell, Gallo \& Fender 2013 with Narayan \& McClintock (2012), and
in SMBHs compare Sikora, Stawarz \& Lasota (2007) with Broderick \&
Fender (2011).

By contrast, the radiation emitted from the jet is fairly well
understood, with spectra separating the Blazars into two types: BL
Lacs and Flat Spectrum Radio Quasars (FSRQs). BL Lacs are typically
completely dominated by the jet emission, emitting a double humped
synchrotron self-Compton (SSC) spectrum. The FSRQs are more complex,
showing clear signatures of a 'normal' AGN disc and broad line region
(BLR), unlike the BL Lacs which generally show no broad lines or disc
emission. This lack of a standard disc/BLR in the BL Lacs is not just
an effect from the relativistically boosted jet emission drowning out
these components: the FSRQs have similarly boosted jet emission yet the
disc and BLR are still clearly visible. Additionally, the presence of
the disc and BLR in FSRQs means that there is an additional source of
seed photons for cooling of relativistic particles in the jet, so
their jet emission includes both SSC and external Compton (EC)
components (Dermer, Schlickeiser \& Mastichiadis 1992; Sikora,
Begelman \& Rees 1994), clearly contrasting with the SSC only jets in BL
Lacs. 

Thus the nature of the accretion flow itself is different in BL Lacs
and FSRQs, with the latter showing a standard disc which is absent from
the former. This can be linked to the clear distinction in Eddington
ratio between BL Lacs and FSRQs, with the BL Lacs all consistent with
$\dot{m} = \dot{M}/\dot{M}_{Edd}< 0.01$ (where $\eta \dot{M}_{Edd}
c^2=L_{Edd}$ and efficiency $\eta$ depends on black hole spin) while
the FSRQs have $\dot{m} > 0.01$ (see e.g.  Ghisellini et al 2010,
hereafter G10). This observed transition in accretion flow properties
occurs very close to the maximum luminosity of the alternative
(non-disc) solutions of the accretion flow equations. These result in
a radiatively inefficient, hot, optically thin, geometrically thick
flow (e.g. ADAFs: Narayan \& Yi 1995) instead of a standard
Shakura-Sunyaev, cool, optically thick, geometrically thin disc. Thus
the BL Lacs have luminosity below this transition and can be
associated with the radiatively inefficient flows, while FSRQs accrete
at higher rates and have standard disks (see e.g.  Ghisellini \&
Tavecchio 2008; G10; Ghisellini et al
2011, Best \& Heckman 2012).

A similar transition is probably present in all types of AGN, not just
the radio loud objects. This predicts that a UV bright accretion disk
is only present at $\dot{m}>0.01$. Most (all?)  radio quiet Seyferts
and Quasars accrete above this limit (see e.g. Woo \& Urry 2002).
Strong UV is required to excite the broad emission lines, and the
material which makes the broad line region (BLR) may also be a wind
from the disk (e.g. Murray \& Chiang 1997, Czerny \& Hryniewicz 2012),
so when the disk is replaced by a hot flow there is no strong UV
emission and no broad lines - hence perhaps the LINERs which are
associated with $\dot{m}<0.01$ (e.g. Satyapal et al 2005).  Additional
evidence for this accretion flow transition comes from the much lower
mass BHBs, which show a dramatic spectral switch around $\dot{m}\sim
0.01$ (Narayan \& Yi 1995; Esin et al 1997 see e.g. Done, Gierlinski
\& Kubota 2007 for a discussion of how this can explain the observed
behaviour of BHBs).

Since all BL Lacs are associated with a low $\dot{m}$ accretion flow,
we test here the hypothesis that all low $\dot{m}$ flows can launch a
jet whose properties are determined simply by mass and mass accretion
rate. We use the simplest possible scalings for how the jet (emission
region size, magnetic field and injected power) scales with these
parameters (Heinz \& Sunyaev 2003; Heinz 2004), anchoring our
scalings onto the fits to individual BL Lac objects of G10. We then
can predict the jet spectra from black holes at any mass and mass
accretion rate. We restrict our work to systems with $\dot{m}<0.01$,
ie. BL Lac type jets, to avoid the additional uncertainties of
external seed photon density scaling with $M$ and $\dot{m}$. We use
cosmological simulations to predict the number densities of black
holes with $\dot{m}<0.01$, and assume that each of these will produce
an appropriately scaled BL Lac type jet. We include electron cooling,
losses due to pair production on extragalactic background light and
orientation effects from beaming to predict how many BL Lacs should be
detected by Fermi. This is a statistical approach to constraining jet
power in the population as a whole (cf. Mart{\'{\i}}nez-Sansigre \&
Rawlings 2011), contrasting with most other previous approaches which
determine jet power from detailed spectral fitting to individual
sources.

We compare our predicted mass and redshift distributions with
observations, and find we overpredict the observed number of BL Lacs
by a factor of 1000. This strongly argues for another parameter apart
from mass and mass accretion rate being required to produce BL Lac
type jets. The most plausible additional factor which can affect the
small size scales of the Fermi emission region is black hole spin. If
this is indeed the answer - and there is longstanding speculation that
high spin is required to produce the most relativistic jets - then
this requires that high spin objects are rare. This is not the case if
the SMBH grows in mass in prolonged accretion episodes, as the
accreted angular momentum can quickly spin the black hole up to
maximal (Volonteri et al 2005; 2007; 2012: Fanidakis et al
2011). Instead, it requires that the SMBH mass build up via accretion
is from multiple, randomly aligned smaller episodes, resulting in low
spin (King et al 2008). The only high spin black holes in these
simulations are the most massive, where the last increase in mass was
via a black hole-black hole coalescence following a major host galaxy
merger (Volonteri et al 2005; 2007; 2012: Fanidakis et al 2011; 2012)

\section{Synchrotron Self-Compton Jets}

We adopt a single zone SSC model of the type used by Ghisellini \&
Tavecchio (2009), which self consistently determines the electron
distribution from cooling. We briefly summarise our model here, with
full details in the Appendix.

We assume a spherical emission region of radius $R$. We
neglect the contribution from regions further out along the jet, as
these only make a difference to the low energy (predominantly radio)
emission. We assume material in the jet moves at a constant bulk
Lorentz factor ($\Gamma$), and that a fraction of the resulting jet
power is used to accelerate electrons in the emission region. The
power injected into relativistic electrons is then $P_{rel}=4/3\pi R^3
\int\gamma m_ec^2 Q(\gamma) d\gamma$, where the accelerated electron 
distribution is a broken power law of the form:

\begin{multline}
Q(\gamma)=Q_0\left(\frac{\gamma}{\gamma_b}\right)^{-n_1}\left(1+\gamma/\gamma_b\right)^{n1-n2} \\ 
\mbox{ for } \gamma_{min}<\gamma<\gamma_{max}
\end{multline} 

These electrons cool by emitting self absorbed synchrotron and
synchrotron self-Compton radiation, so the seed photon energy density
$U_{seed}=U_B+g(\gamma)U_{sync}$ includes both the magnetic energy
density $U_B=B^2/8\pi$, and the fraction $g(\gamma)$ of the energy
density of synchrotron seed photons, $U_{sync}$, which can be Compton
scattered by electrons of energy $\gamma$ within the Klein-Nishina
limit. This gives rise to a steady state electron distribution,
$N(\gamma)=-\dot{\gamma}^{-1}\int_\gamma^{\gamma_{max}} Q(\gamma')d\gamma'
$, where the rate at which an electron loses energy $\dot{\gamma}
m_ec^2=4/3\gamma^2 \sigma_T c U_{seed}$. However, this assumes
that the electrons can cool within a light crossing time, but the
cooling timescale $t_{cool}=\gamma/\dot{\gamma}$ itself
depends on $\gamma$, with high energy electrons cooling fastest.  We
calculate the Lorentz factor that can just cool in a light crossing
time of the region,$\gamma_{cool}$, and join smoothly onto the
accelerated electron distribution below this. The full self consistent
electron distribution can be characterised by $N(\gamma)=Kn(\gamma)$,
where $K$ is the number density of electrons at $\gamma=1$ and
$n(\gamma)$ incorporates all the spectral shape.
We calculate the resulting (self absorbed) synchrotron and self
Compton emission using the delta function approximation as this is
much faster than using the full kernel but is accurate enough for our
statistical analysis (Dermer \& Menon 2009).

This jet frame emission is boosted by the bulk motion of the jet,
with the amount of boosting depending on both $\Gamma$ and the
orientation of the jet. The emission is then 
cosmologically redshifted and attenuated due to pair production on the
extragalactic infrared background light (though this is generally small
for the Fermi bandpass) to produce the observed flux. 

The parameters of our model are therefore: 
\begin{itemize}
\item
Physical parameters of the jet: $\Gamma$ and radius of emission region $R$.
\item
The magnetic field of the emission region and power injected into relativistic electrons ($B$ and $P_{rel}$).
\item
Parameters of the injected electron distribution: $\gamma_{min}$, $\gamma_b$, $\gamma_{max}$, $n_1$ and $n_2$.
\end{itemize}

We adopt the cosmology used in the Millennium simulations: $h=0.72,$ $\Omega_m=0.25$, $\Omega_{vac}=0.75$ (Springel et al 2005; Fanidakis et al 2011).

\section{Scaling Jets}

We assume that the acceleration mechanism is the same for all BL Lacs,
giving the same injected electron distribution, regardless of mass and
accretion rate. We also assume all jets are produced with the same
$\Gamma$. This leaves three remaining parameters: $R$, $B$ and $P_{rel}$.

We scale $R\propto M$, since all size scales should scale with the
mass of the black hole (Heinz \& Sunyaev 2003). We assume the jet
power is a constant fraction of the total accretion power,
$P_j\propto\dot{m}M$. This assumption is valid whether the jet is
powered by the accretion flow or the spin energy of the black hole,
since extraction of black hole spin energy relies on magnetic fields
generated in the accretion flow, which will be affected by accretion
rate. A constant fraction of the total jet power is then injected into
relativistic particles and magnetic fields. Hence $P_{rel}\propto
P_j\propto\dot{m}M$.

Energy density in the jet frame is related to power in the rest frame
via $P=\pi R^2 \Gamma c U$, so $P_B\propto R^2 U_B\propto\dot{m}M$,
hence $B\propto U_B^{1/2}\propto (\dot{m}/M)^{1/2}$. Therefore all
energy densities should scale as $U_B\propto U_{rel}\propto
(\dot{m}/M)^{1/2}$.

We anchor this with parameters from the fit to the classic low peaked BL
Lac (LBL) object, 1749+096 from G10, which is relatively near to the
top of the BL Lac accretion rate range. This gives 
$M_0 = 7\times10^8$, $R_0 = 172\times10^{15} cm$, $B_0 =
1G$, $P_{rel,0}=3.5\times10^{42} erg s^{-1}$, $\Gamma=15$,
$\gamma_{min}=1$, $\gamma_b=2\times10^3$, $\gamma_{max}=1\times10^5$,
$n_1=0.9$, $n_2=2.8$, and we scale $R$, $P_{rel}$ and $B$ as:

\begin{equation}
R = R_0\frac{M}{M_0}
\end{equation}
\begin{equation}
P_{rel} = P_{rel,0}\frac{\dot{m}}{\dot{m_0}}\frac{M}{M_0} 
\end{equation}
\begin{equation}
B = B_0\left(\frac{\dot{m}}{\dot{m_0}}\frac{M_0}{M}\right)^{1/2}
\end{equation}

We calculate the accretion rate of 1749+096 following the method of
G10. Assuming the jet is maximal ($P_j=\dot{M}c^2$), we sum the power
in magnetic fields, relativistic electrons and the bulk motion of cold
protons to calculate $P_j$, giving:

\begin{equation}
\dot{m_0} = \frac{P_j\eta}{1.38\times10^{38}(M/M_{\odot})}\sim 3.5 \times10^{-3}
\end{equation}

For their value of $\eta=0.08$. Fig 1 shows our model spectrum,
together with the model and data from G10. The two models differ
slightly due to our use of the delta function approximation to speed
up calculation time. Nevertheless the two models are in agreement
within $\sim 0.3$ dex, and crucially our model reproduces the correct
level of Fermi flux (red bow tie).

\begin{figure} 
\centering
\begin{tabular}{l}
\leavevmode  
\epsfxsize=8cm \epsfbox{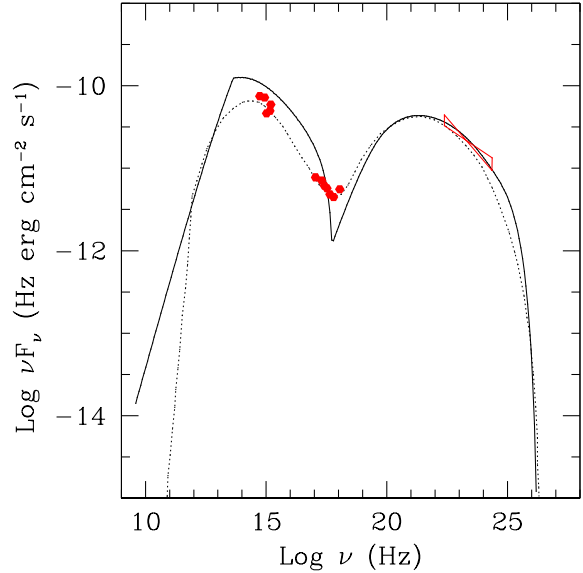}\\
\end{tabular}
\caption{Model spectrum using parameters for 1749+096 (z=0.322) from Ghisellini et al (2010) (solid black line). Dotted line shows their spectrum for the same parameters and red points show their data.}
\label{fig1}
\end{figure}

\section{Transition from High Frequency Peaked to Low Frequency Peaked BL Lacs with Accretion Rate}

\begin{figure*} 
\centering
\begin{tabular}{l|c|r}
\leavevmode  
\epsfxsize=5cm \epsfbox{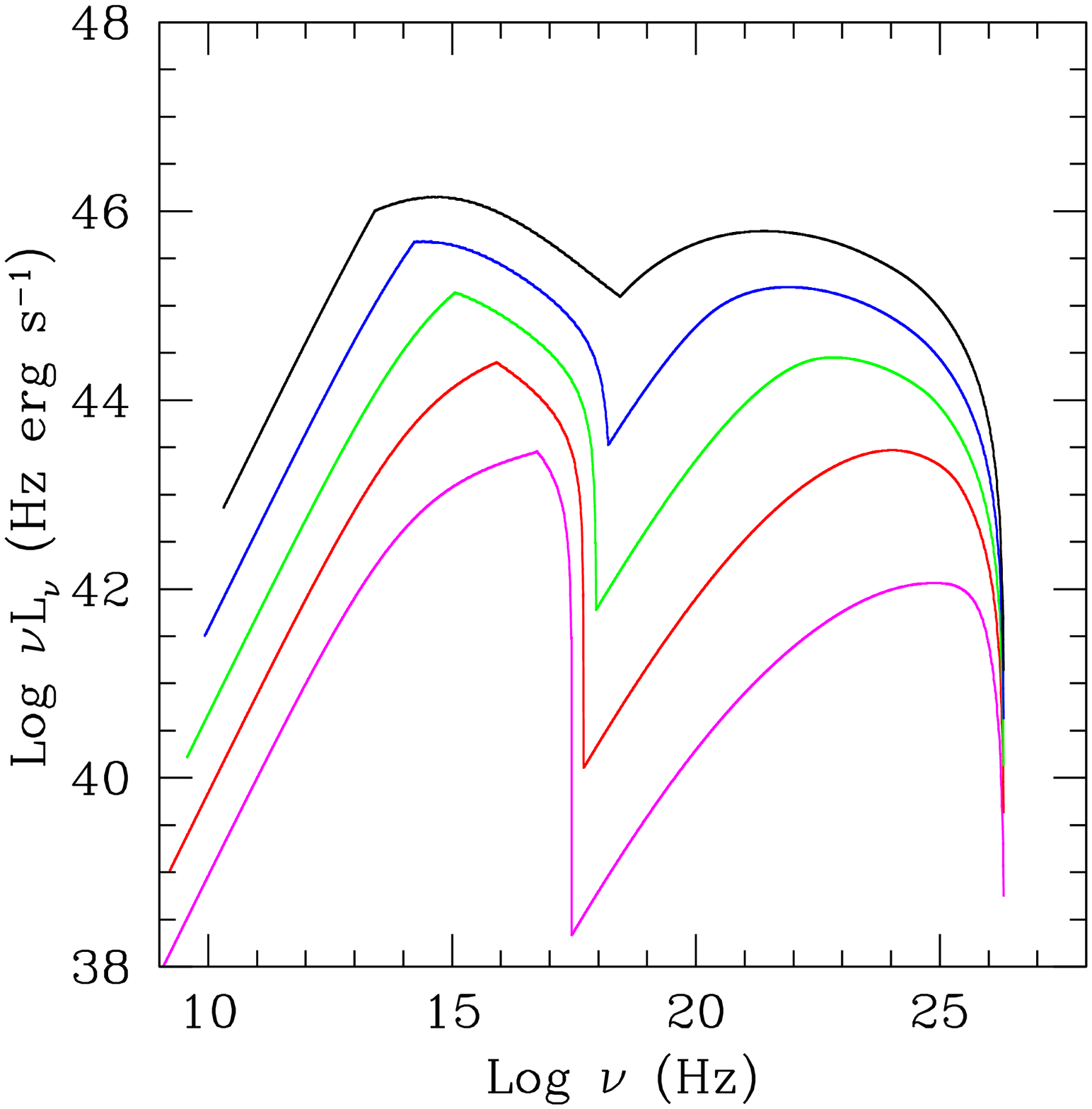} &
\epsfxsize=5cm \epsfbox{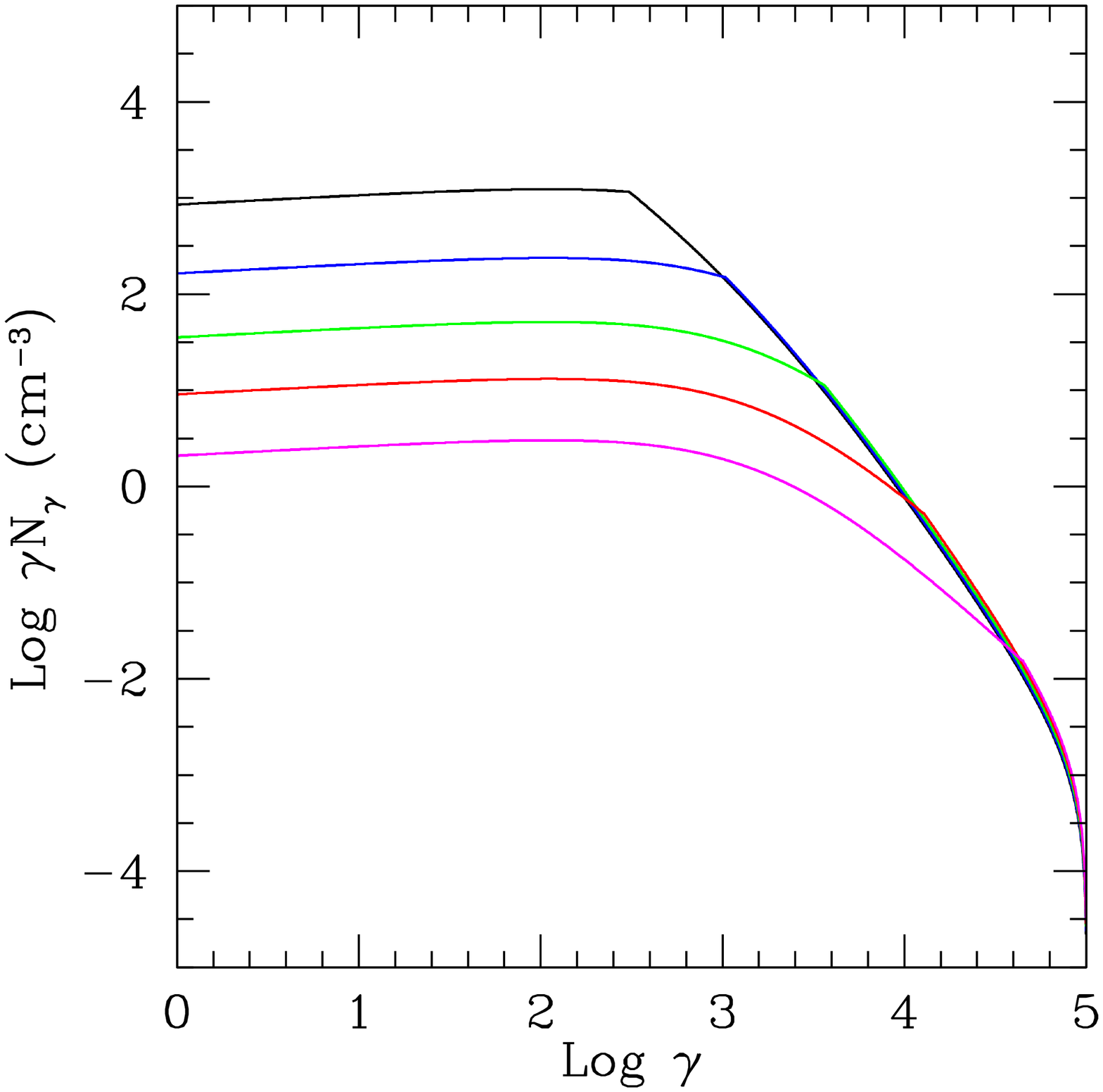} &
\epsfxsize=5cm \epsfbox{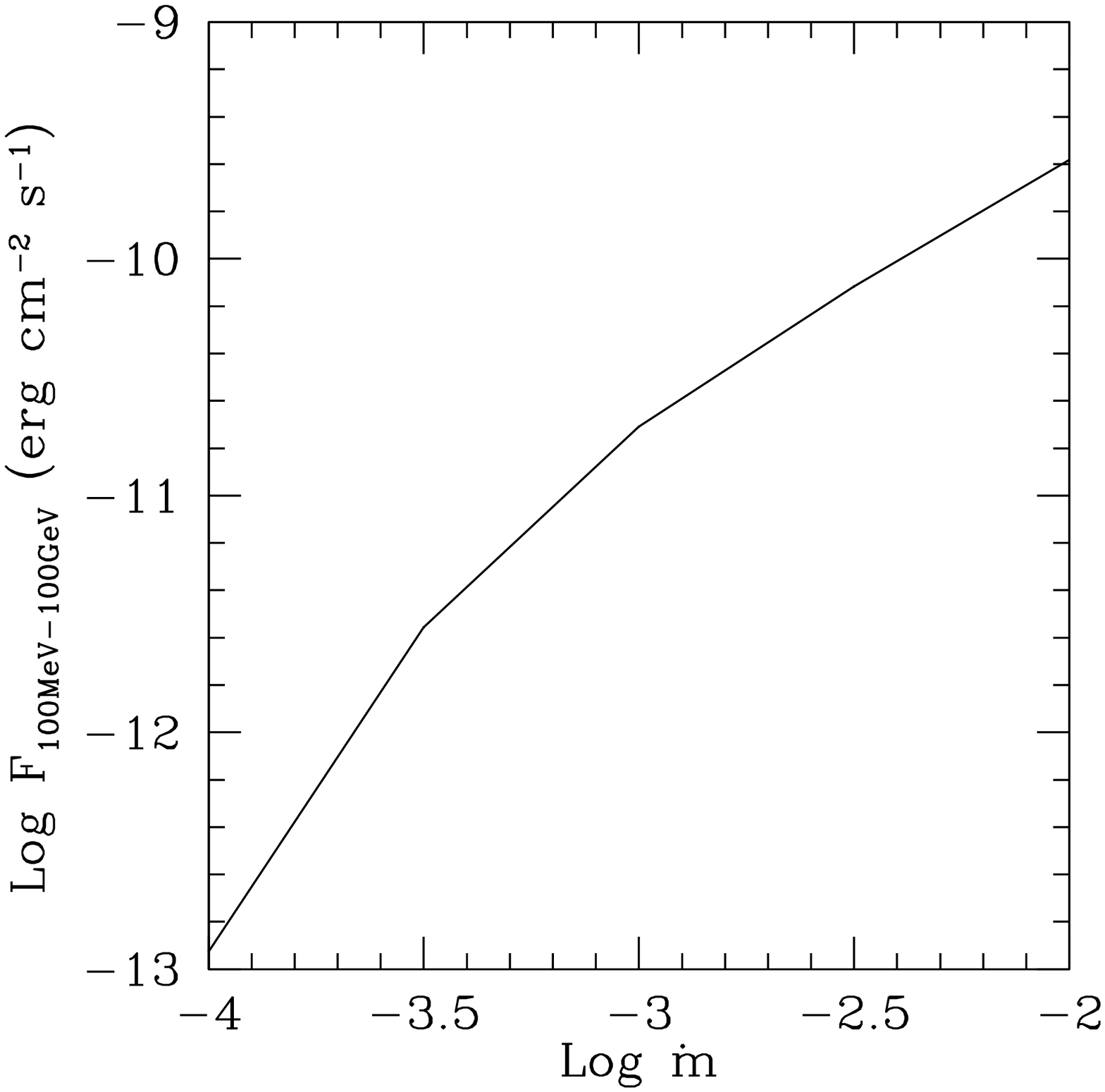}\\
\end{tabular}
\caption{a). BL Lac model SEDs for fixed black hole mass and increasing accretion rate ($\dot{m}=10^{-4}$ (magenta), $3\times 10^{-4}$ (red), $10^{-3}$ (green), $3\times 10^{-3}$ (blue) and $10^{-2}$ (black), $M_{BH}=10^8 M_\odot$). b). Corresponding steady state electron distributions. c). Fermi flux as a function of accretion rate, using mass and distance of model spectrum.}
\label{fig1}
\end{figure*}

\begin{figure*} 
\centering
\begin{tabular}{l|c|r}
\leavevmode  
\epsfxsize=5cm \epsfbox{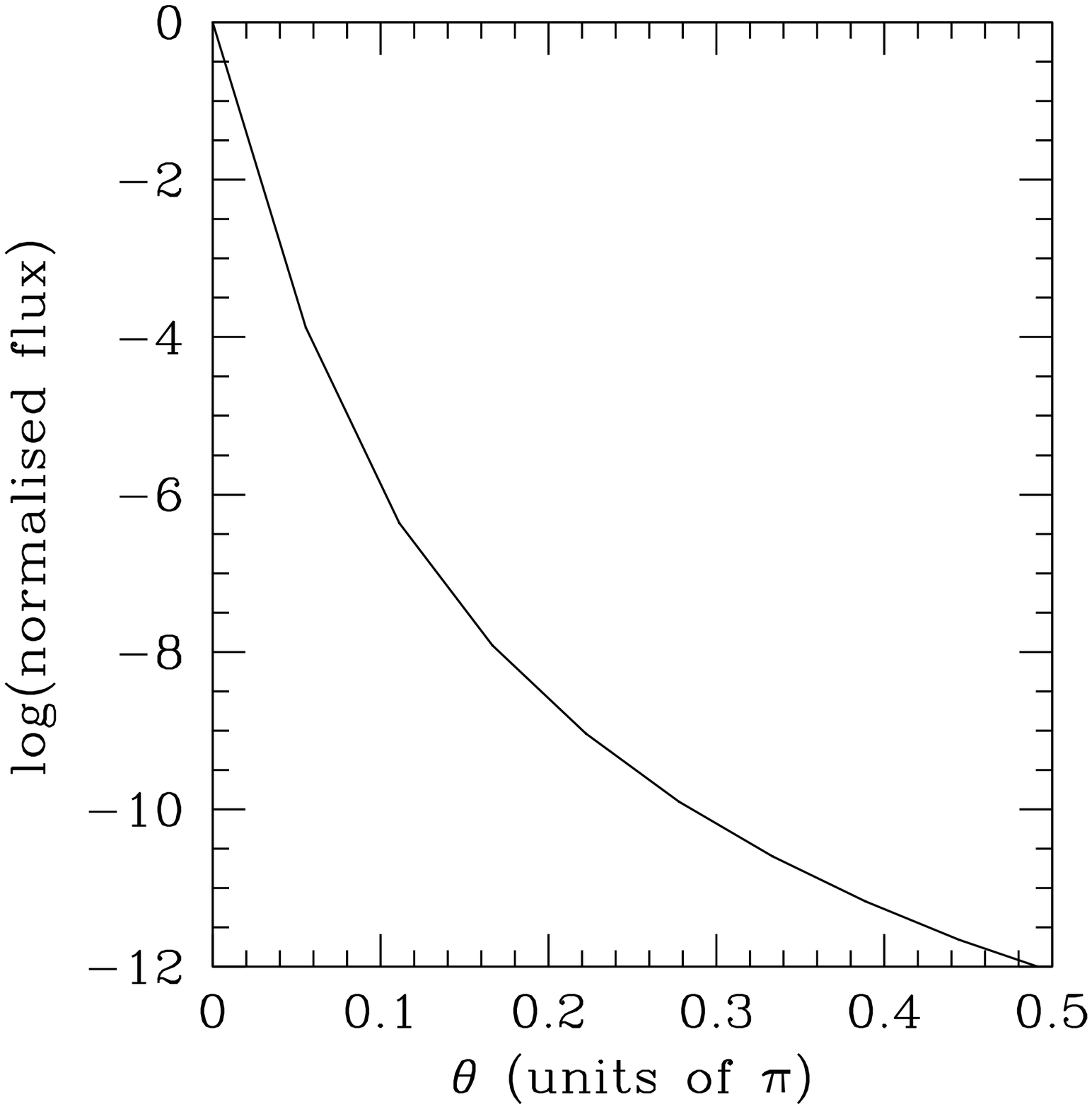} &
\epsfxsize=5cm \epsfbox{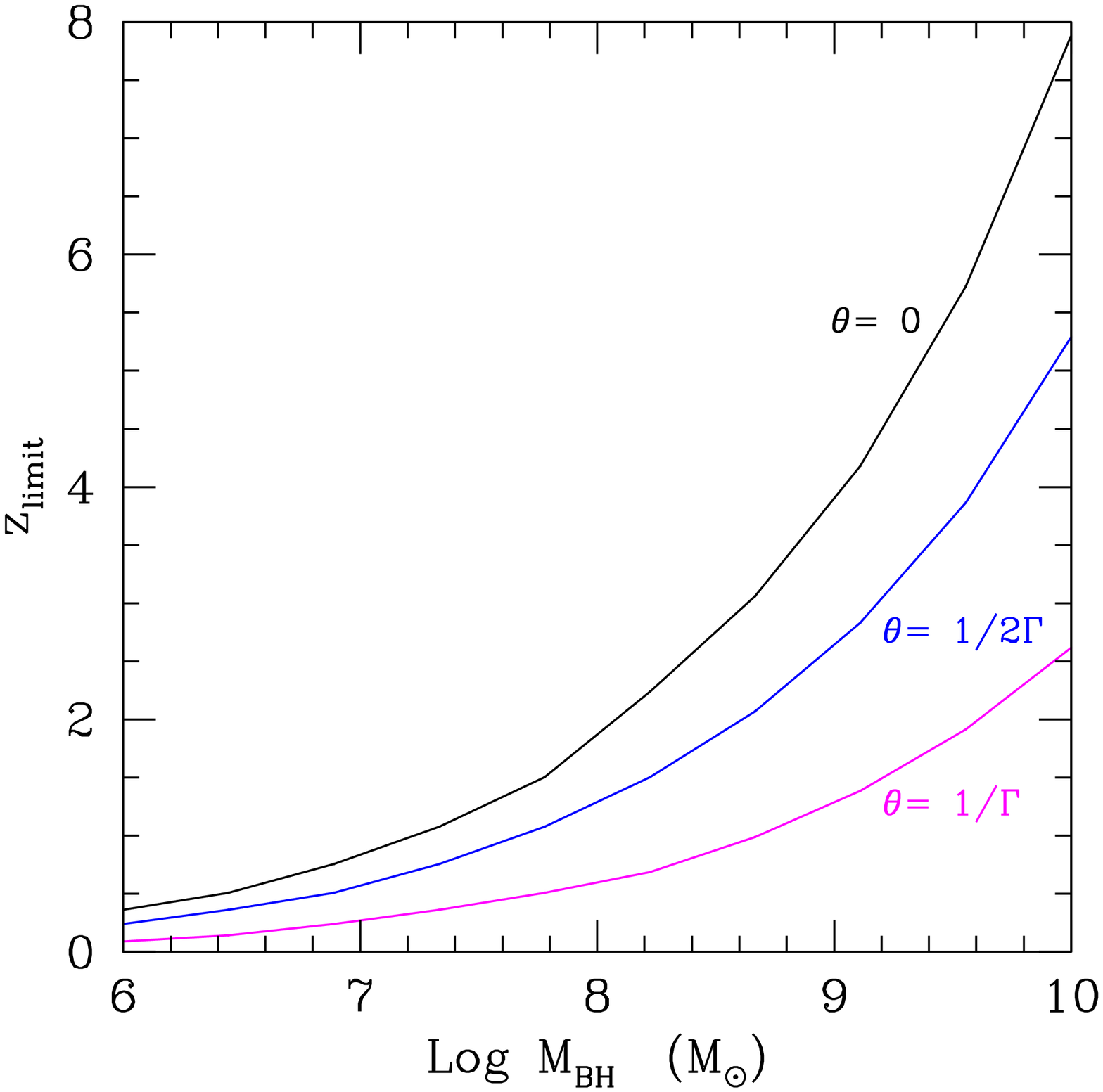} &
\epsfxsize=5cm \epsfbox{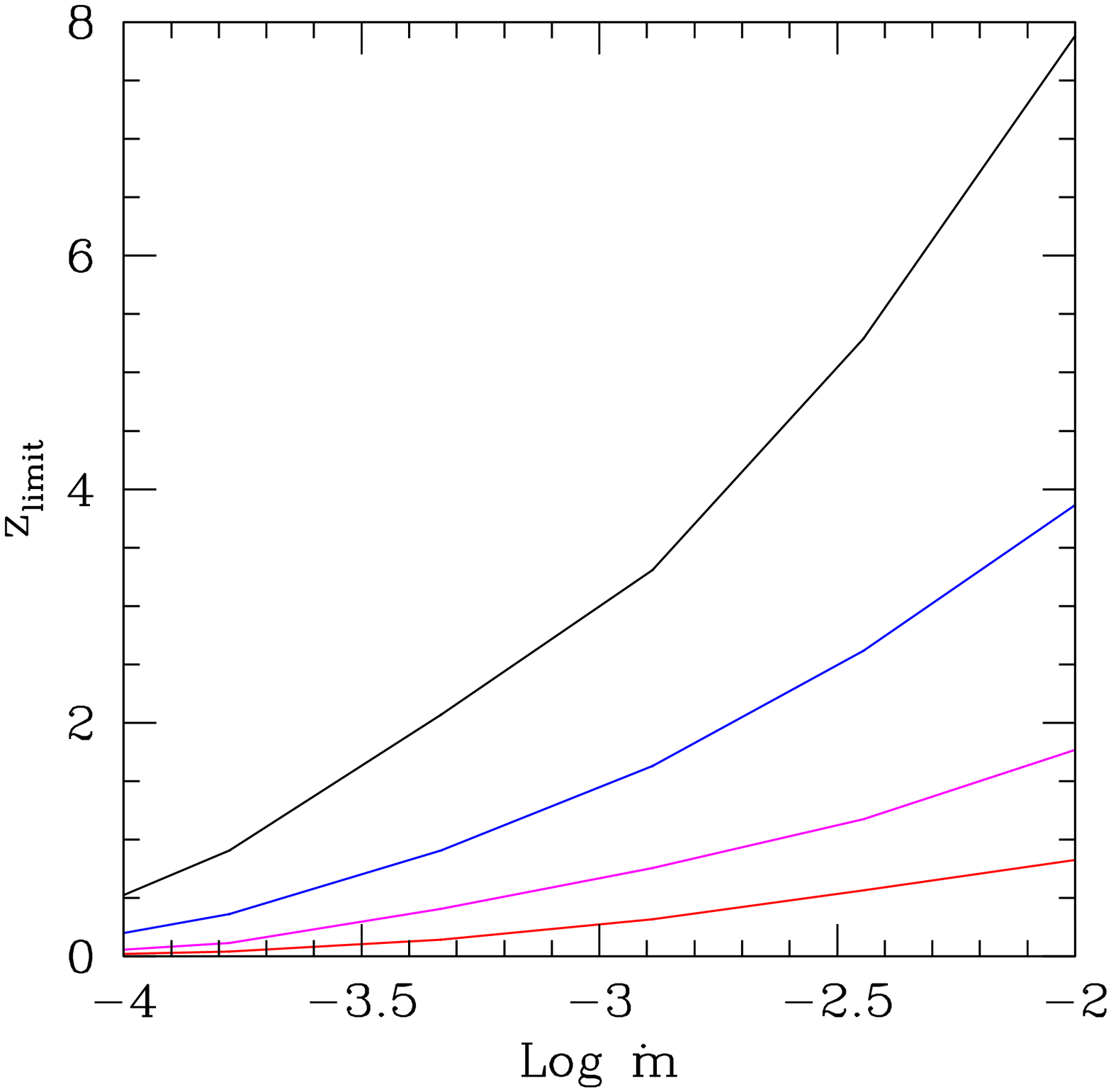} \\
\end{tabular}
\caption{a). Decrease in observed flux with increasing viewing angle, where $\theta$ is measured in radians from the jet axis, for $\Gamma=15$. b). Redshift limits for Fermi visible BL Lacs as a function of black hole mass, for increasing viewing angle and $\dot{m}=10^{-2}$. c). Redshift limits for Fermi visible BL Lacs as a function of accretion rate, for $M_{BH}=10^7$ (red), $10^8$ (magenta), $10^9$ (blue) and $10^{10} M_{\odot}$ (black) and $\theta=0$.}
\label{fig1}
\end{figure*}

We limit our model to a maximum accretion rate of $\dot{m}=10^{-2}$,
since above this the accretion flow is expected to make a transition
to a radiatively efficient thin disc. The strong UV and consequent
broad line region emission provide additional seed photons, switching
the main Fermi radiation process from SSC (BL Lacs) to EC (FSRQs). 

Fig 2a shows a sequence of spectra with $R$, $P_{rel}$ and $B$
scaling as described above for $\dot{m}=10^{-2}$ (black) - $10^{-4}$
(magenta) for constant mass. This shows the systematic decrease in
luminosity, coupled to a change in spectral shape from a low
synchrotron peak energy (optical: LBL) to a high synchrotron peak
energy (X-ray: HBL) as shown by Ghisellini \& Tavecchio (2008; 2009). 

We can compare the Fermi flux levels of our lower accretion rate
spectra with observed HBLs. The HBL 1959+650 (see Tavecchio et al 2010
for a spectrum, G10 for spectral fitting parameters) has a mass of
$2\times10^8M_{\odot}$, so only slightly larger than the
$10^8M_{\odot}$ system shown in Fig 2a. 1959+650 has an injected
$P_{rel}$ of $7\times10^{40} ergs^{-1}$ (G10), corresponding in our
scalings to $\dot{m}=2.45\times10^{-4}$. So it should have a similar
Fermi flux to the red spectrum of fig 2a, which corresponds to
$\dot{m}=3\times10^{-4}$, $M=10^8$. The observed $\log(\nu L(\nu))$
Fermi flux of 1959+650 at $10^{23}Hz$ is $~43.5 Hz ergs^{-1}$, which
is consistent with our red spectrum.

The changing shape of the emitted spectrum with accretion rate is due to the
decrease in seed photons for electron cooling at lower $\dot{m}$, as
shown explicitly by the the corresponding self consistent electron
distributions in Fig 2b. The Lorentz factor of electrons which can
cool in a light crossing time is $\gamma_{cool}\propto 1/(R
U_{seed})\propto (1/M) (M/\dot{m})\propto 1/\dot{m}$. 
The lowest mass accretion rate ($\dot{m}=10^{-4}$, magenta)
shows cooling only for the highest Lorentz factors, with
$\gamma_{cool}\sim 10^{4.5}$. Below this the shape of the electron
distribution is the same as the injected distribution, with a smooth
break at $\gamma_b\sim 10^3$. As $\dot{m}$ increases, the sharp break
at $\gamma_{cool}$ moves to lower Lorentz factors. For the
black electron distribution corresponding to $\dot{m}=10^{-2}$,
$\gamma_{cool}$ is comparable to $\gamma_b$. This is clear
from the black spectrum in Fig 2a, where the spectral peak is now
produced by the cooled electron distribution above $\gamma_{cool}$.

Increasing cooling, as a result of increasing accretion rate,
therefore provides a natural explanation for the existence of high
frequency peaked (HBL) and low frequency peaked (LBL) BL Lacs
(Ghisellini \& Tavecchio 2008; 2009). In the context of our model,
HBLs correspond to black holes with very low accretion rates. There is
very little cooling and the bulk of the synchrotron emission is
produced by electrons with Lorentz factors close to
$\gamma_{max}$. Their electron distributions most closely resemble the
original injected distributions. LBLs correspond to black holes
with higher accretion rates, where cooling becomes increasingly
important and the bulk of the energy is produced by electrons close to
$\gamma_{cool}$.

Since HBLs are at lower accretion rates they are intrinsically fainter
and so should be observed at lower redshifts than LBLs. This is indeed
observed (Shaw et al 2013). Fermi sensitivity is also a strong
function of spectral index, decreasing with spectral hardness (Nolan
et al 2012). Since LBLs have softer spectra this suggests Fermi will
preferentially select LBLs over HBLs due to spectral shape as well as
flux.

Fig 2a also shows that as $\dot{m}$ increases, the ratio of the
Compton to synchrotron luminosities changes. With our scalings,
$L_{sync}\propto R^3 U_B K$ and $L_{comp}\propto R^3U_{sync}K\propto
R^3(RU_BK)K$, ie. $L_{sync}/L_{comp}\propto 1/(RK)$, where $K$ is the
normalisation of the steady state electron distribution. If there is
complete cooling, ie. $\gamma_{cool}<\gamma_{min}$, then $K\propto
Q_0/U_{seed}\propto (\dot{m}/M^2)(\dot{m}/M)^{-1}\propto 1/M$ which is
independent of accretion rate. However the BL Lac spectra do not show
complete cooling (see Fig 2b). If there is no cooling, $K\sim RQ_0/c
\propto \dot{m}/M$. The BL Lac spectra lie in this regime where the
cooling is incomplete, hence $L_{sync}/L_{comp}\propto
1/\dot{m}$. This can be seen in Fig 2b, where the normalisation of the
electron distribution at $\gamma=1$ increases with $\dot{m}$. The
scaling is not exactly $K\propto\dot{m}$, since there is an additional
dependence on $\dot{m}$ introduced by $\gamma_{cool}$ decreasing through
the intermediate regime.

Fig 2c shows how the flux in the Fermi band drops with accretion
rate. For higher accretion rates, cooling is efficient, so
$L_{comp}\propto\dot{m}$. For low accretion rates, cooling is
inefficient so $L_{comp}\propto\dot{m}^3$.

However, a more detailed comparison of Fig 2a to the data 
in the 'blazar sequence' shows evidence that the Compton flux changes 
more slowly with decreasing mass accretion rate due to an increase in
the maximum Lorentz factor of the accelerated electron distribution 
(Ghisellini \& Tavecchio 2008; 2009). Again, this can be a consequence
of the different cooling environment, where electrons are accelerated
to a maximum energy which is set by a balance between the acceleration
timescale and the cooling timescale. Thus the accelerated electron
distribution may itself change with cooling, such that
$\gamma_b\propto \gamma_{max}\propto 1/U_{seed}$. We will consider such
models later in the paper.

\section{BL Lac Visibility}

\begin{figure*} 
\centering
\begin{tabular}{l|c|r}
\leavevmode  
\epsfxsize=5cm \epsfig{file=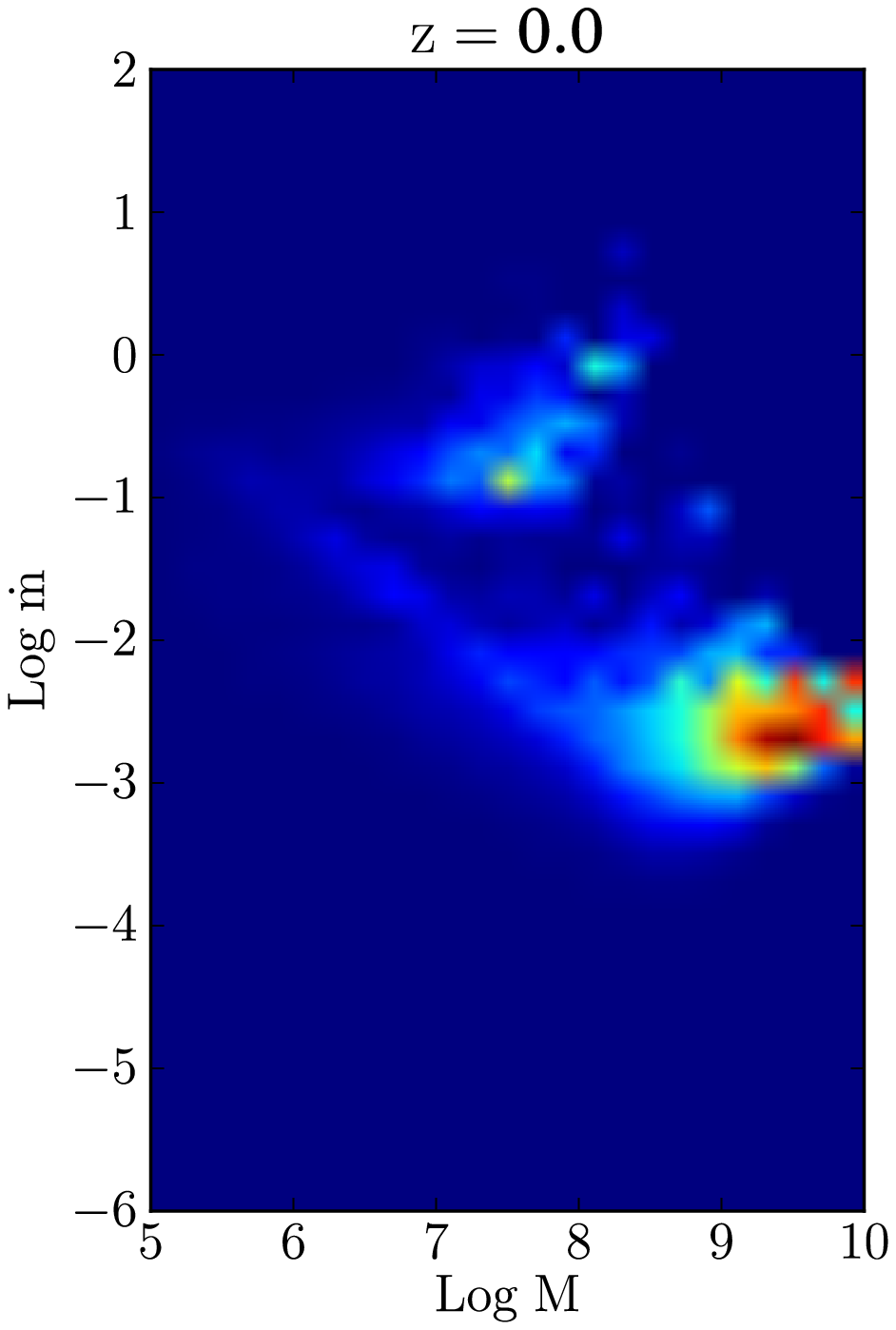,clip=,width=6cm} &
\epsfxsize=5cm \epsfig{file=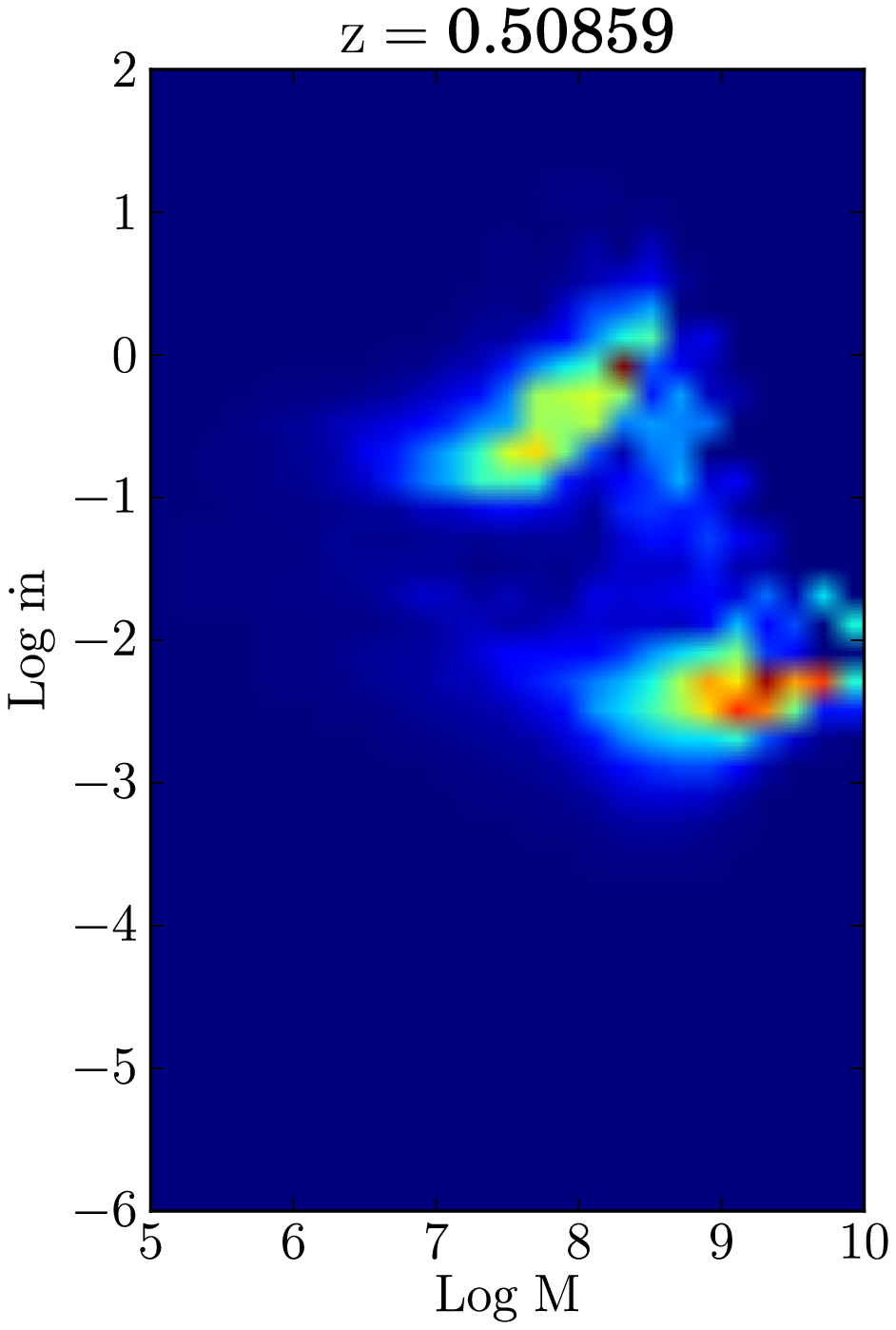,clip=,width=6cm} &
\epsfxsize=5cm \epsfig{file=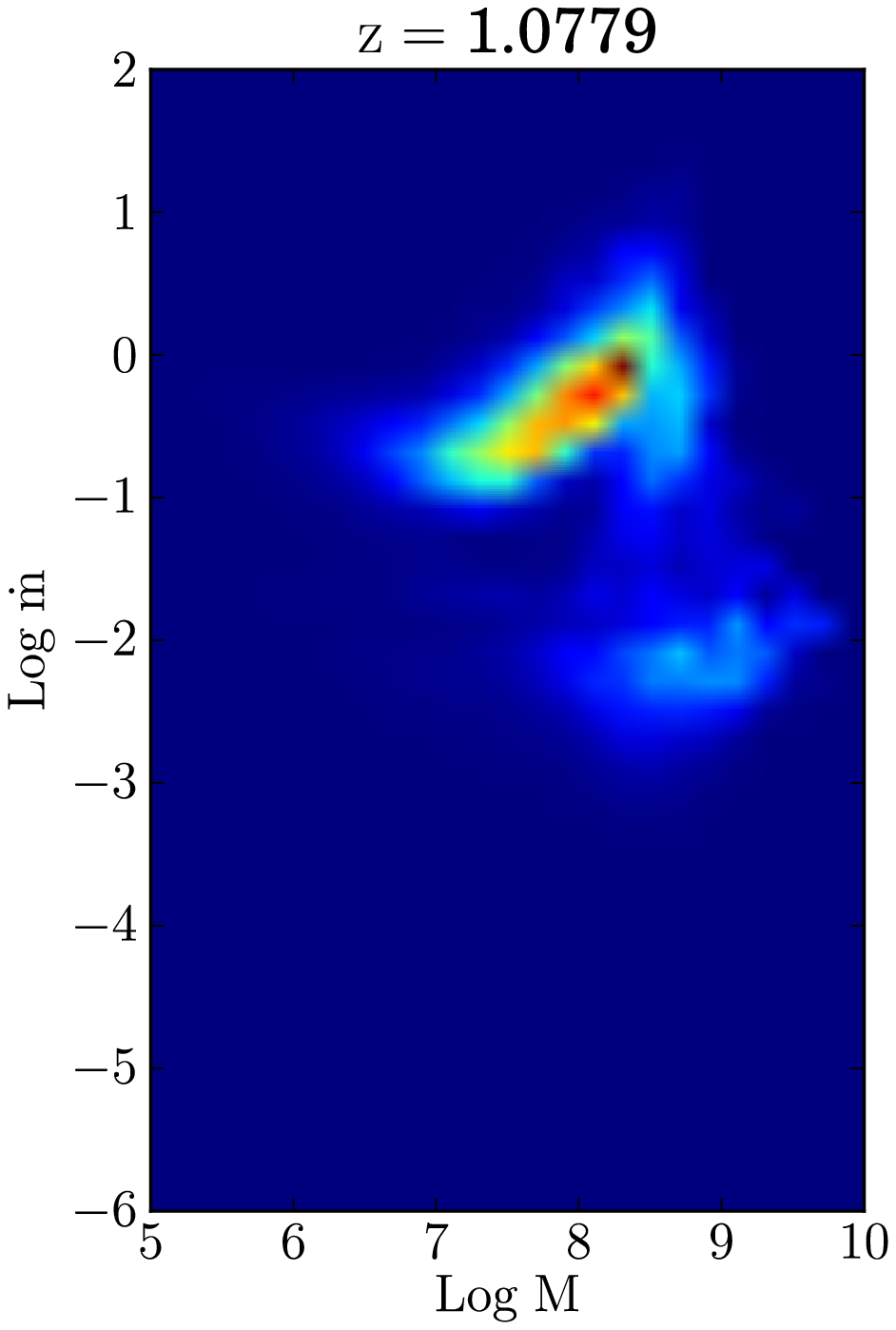,clip=,width=6cm} \\
\epsfxsize=5cm \epsfig{file=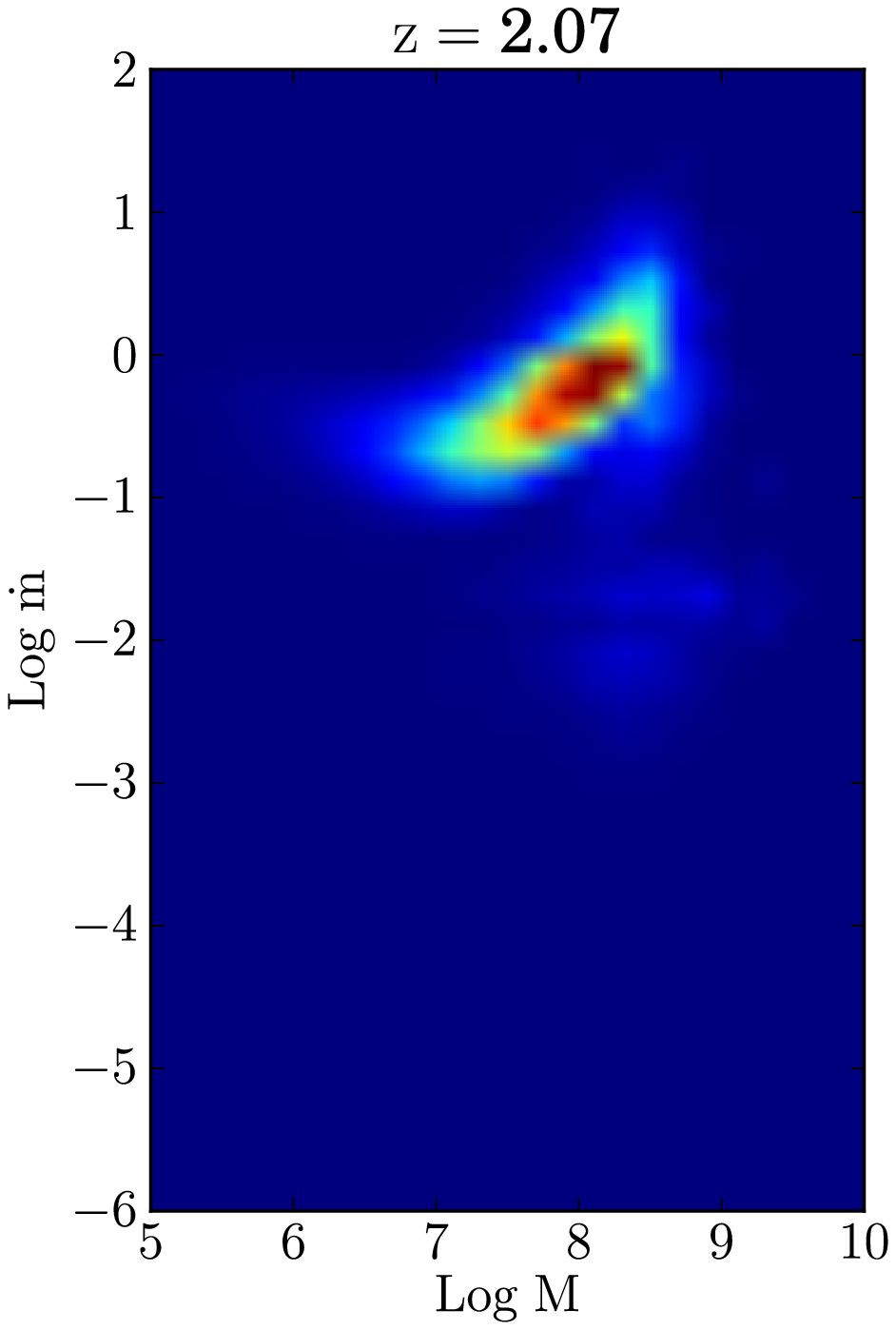,clip=,width=6cm} &
\epsfxsize=5cm \epsfig{file=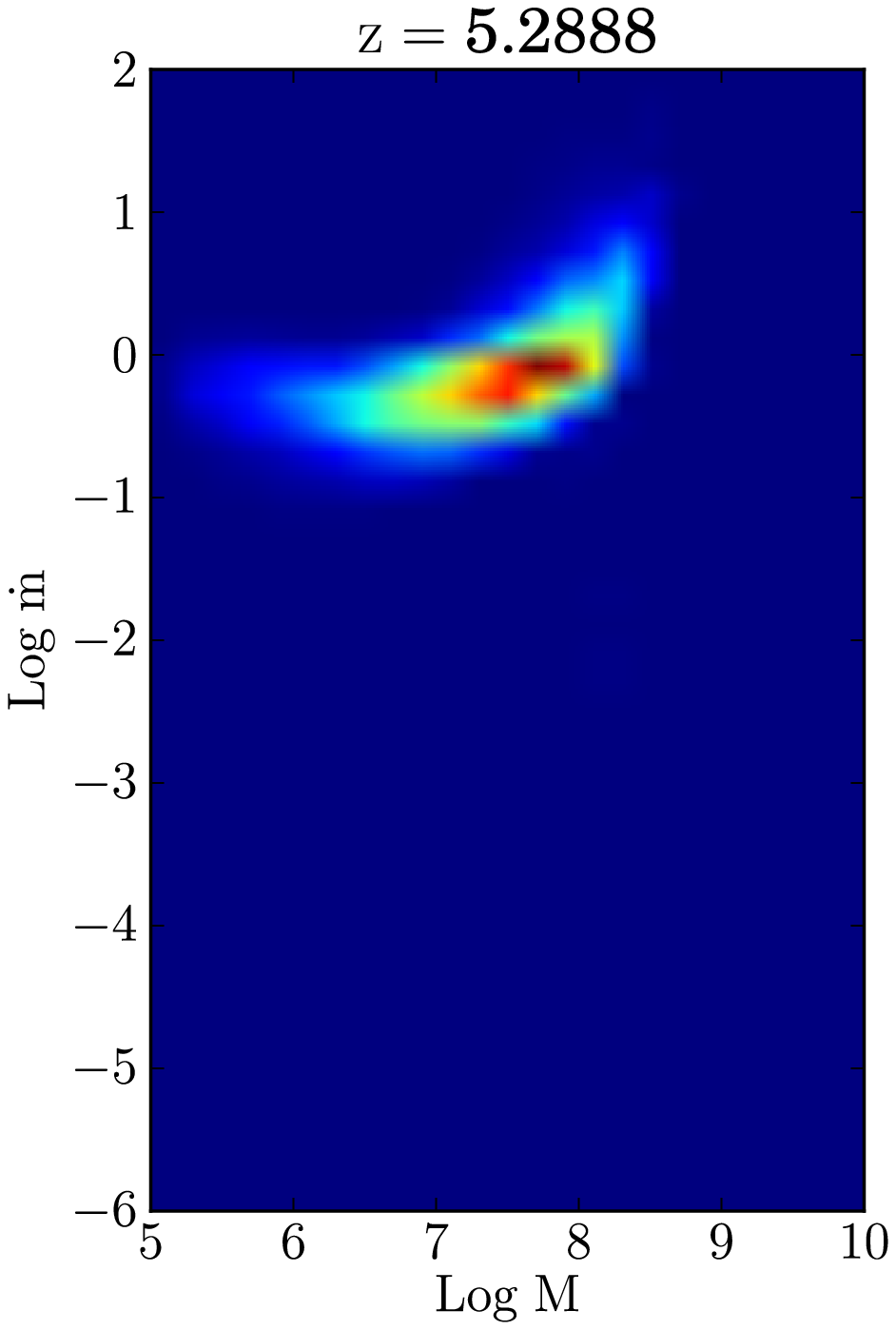,clip=,width=6cm} &
\epsfxsize=5cm \epsfig{file=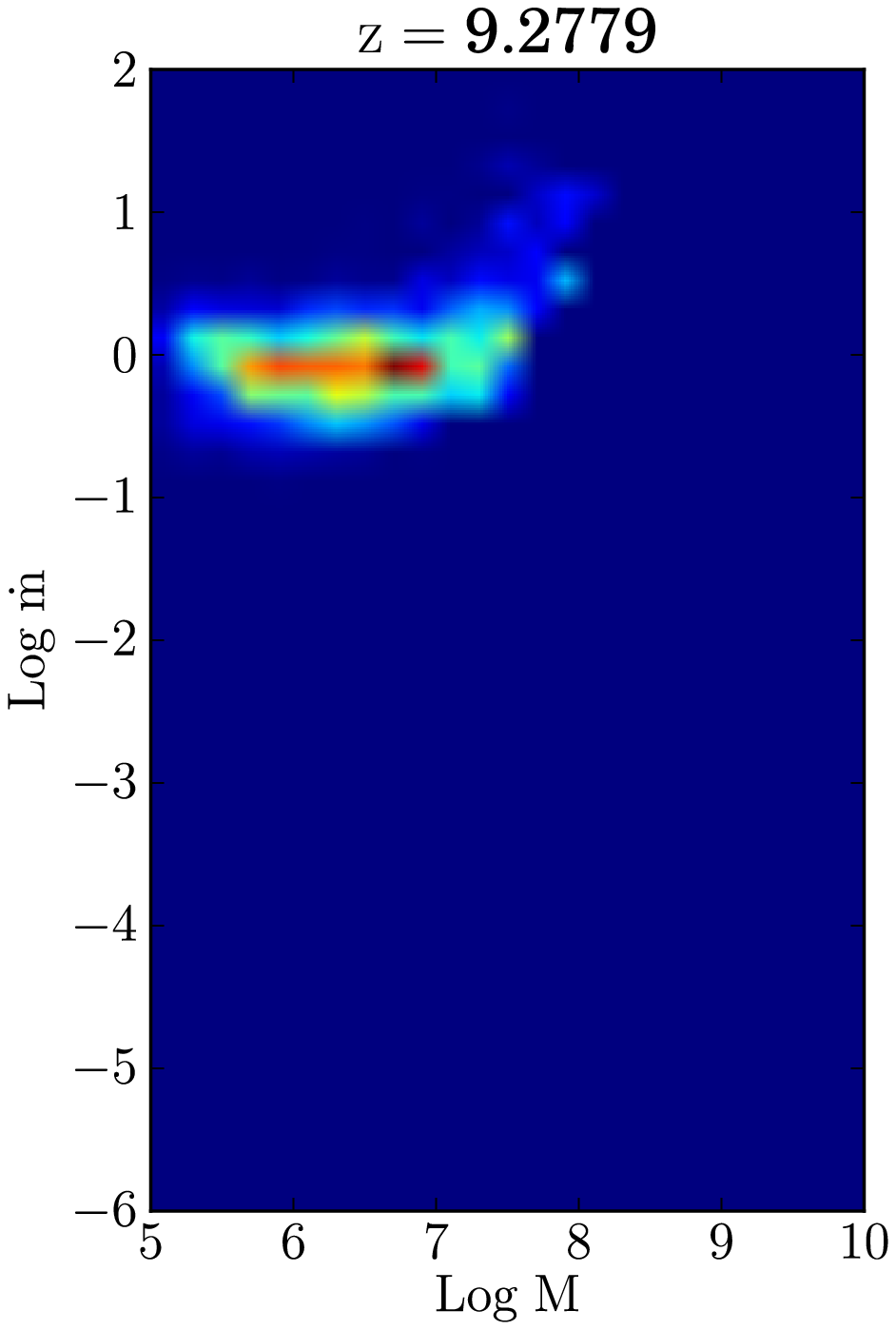,clip=,width=6cm} \\
\end{tabular}
\caption{Predicted mass and accretion rate distribution of accreting black holes at increasing redshift from Millennium simulation. Colours trace luminosity density, with red showing the mass and accretion rates at which the maximum accretion luminosity is emitted at each redshift.}
\label{fig1}
\end{figure*}

The visibility of a BL Lac is strongly affected by viewing angle. Fig
3a shoes how sharply the observed luminosity decreases for our assumed
$\Gamma=15$ with increasing viewing angle, where $\theta$ is measured
in radians from the jet axis. Thus there is a difference of $10^{12}$
between the observed flux from a face on jet compared to an edge on
jet. 

The more distant the source, the more closely aligned to our line of
sight the jet must be in order to boost the observed flux to a visible
level. We define a flux limit of $F_{100MeV-100GeV}>5\times 10^{-12}
erg s^{-1}cm^{-2}$ from the Fermi 2 year catalogue (Nolan et al 2012), and show in Fig 3b
the limiting redshift, $z_{limit}$, at which a BL Lac at
$\dot{m}=10^{-2}$ with different masses can be detected by Fermi
at different inclination angles. We include the effects of absorption
from pair production on the extragalactic IR background using the
model of Kneiske \& Dole (2010), though this is negligible.  Only the
most massive black holes, $\sim 10^{10} M_{\odot}$ which are most
closely aligned to our line of sight can be seen out beyond
$z=4$. $z_{limit}$ drops by a factor of $\sim 3$ if the inclination
angle is increased from $0$ to the more statistically likely
$1/\Gamma$. This represents a change of just $\sim 4^\circ$ for
$\Gamma=15$ used in our calculations. For a typical BL Lac mass
of $10^9 M_{\odot}$ viewed at $1/\Gamma$ the 
maximum observable redshift is $z\sim 1$, increasing to $4$ only for
the most face on jets. 

\begin{figure*} 
\centering
\begin{tabular}{l|c|r}
\leavevmode  
\epsfxsize=5cm \epsfbox{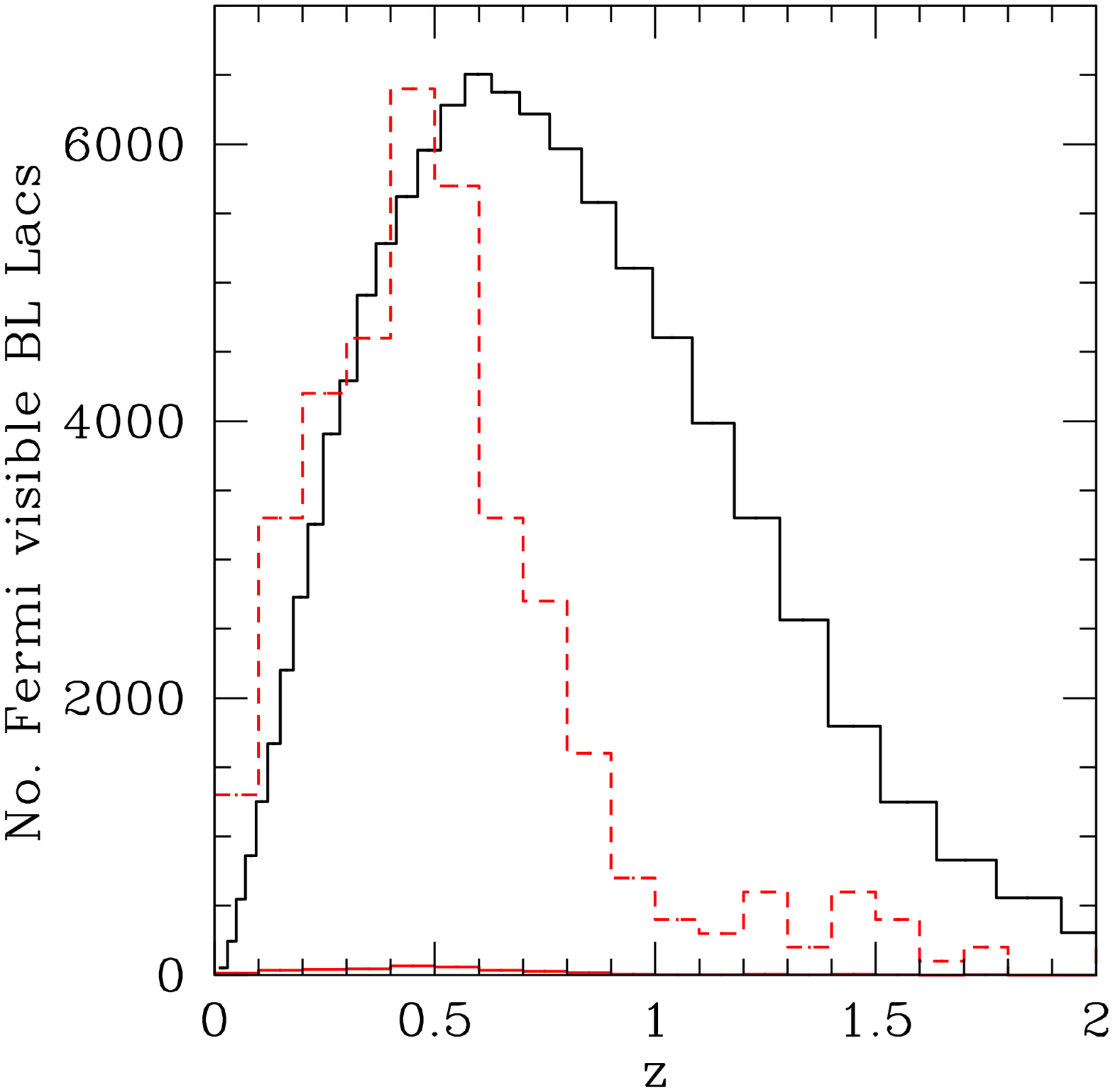} &
\epsfxsize=5cm \epsfbox{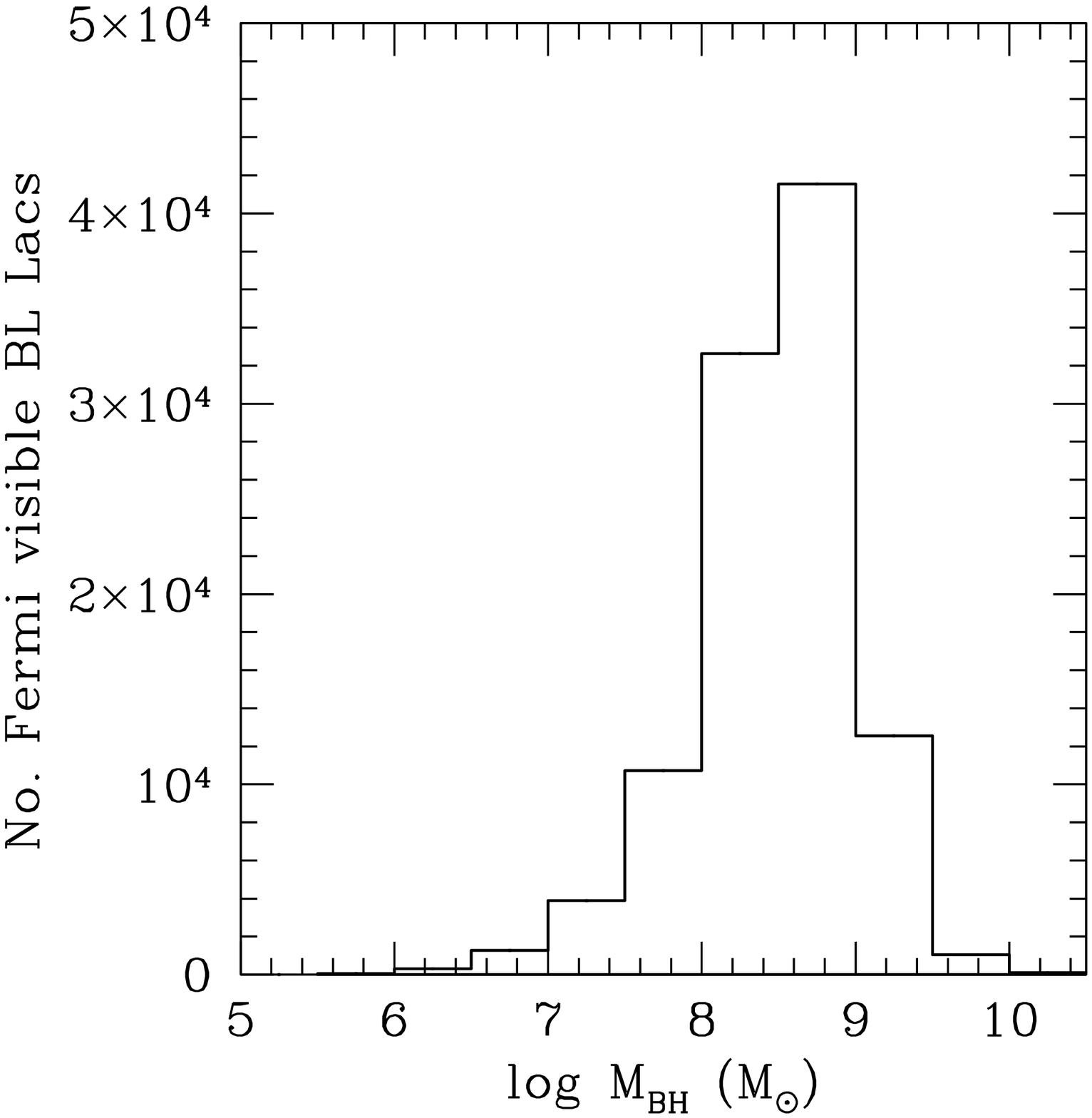} &
\epsfxsize=5cm \epsfbox{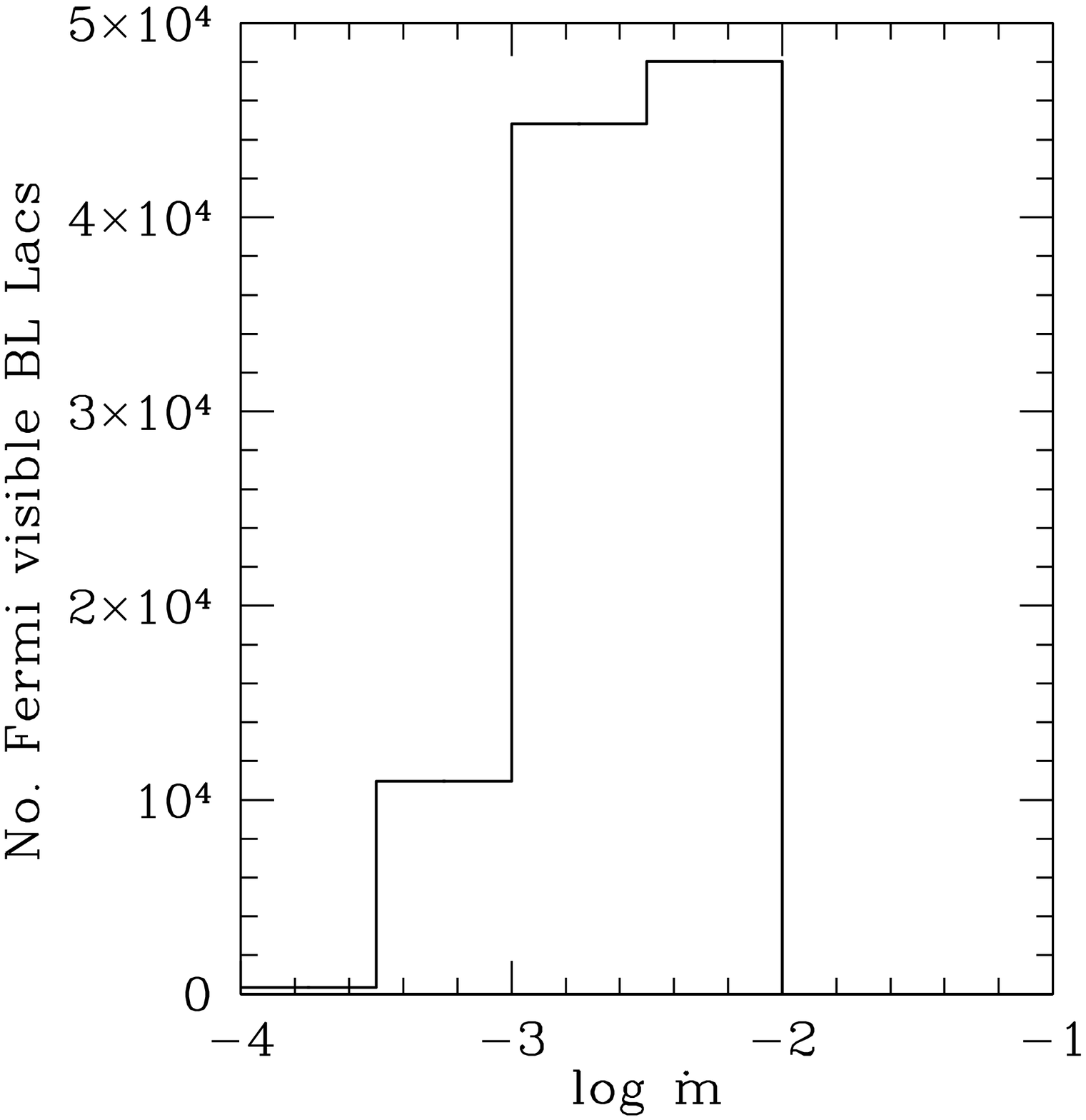} \\
\end{tabular}
\caption{a). Predicted redshift distribution of Fermi visible BL Lacs, assuming BHs of all spins accreting below $\dot{m}=10^{-2}$ produce a BL Lac type jet (black). Red solid line shows observed redshift distribution of Fermi detected BL Lacs. Red dashed line shows observed redshift distribution $\times 100$. b). Predicted mass distribution of Fermi visible BL Lacs. c). Predicted accretion rate distribution of Fermi visible BL Lacs.}
\label{fig1}
\end{figure*}

Fig 3c shows how the redshift limit drops as a function of accretion
rate for each mass ($10^{10}$ (black), $10^9$ (blue), $10^8$ (magenta) and $10^7 M_{\odot}$ (red))
black hole for $\theta=0$. If LBLs correspond to BL Lacs at
$\dot{m}\sim 10^{-2}$ and HBLs at $\dot{m}<10^{-3}$ this shows how the
redshift limits for the two populations should differ, with the
majority of HBLs being observed below $z=3$. Shaw et al (2013) find
this to be the case, with the distribution of LBLs extending to higher
$z$, although they find the means of both populations are well below $z=3$.

\section{Predicted BL Lac Population from Cosmological Simulations}

\begin{figure*} 
\centering
\begin{tabular}{l|c|r}
\leavevmode  
\epsfxsize=5cm \epsfbox{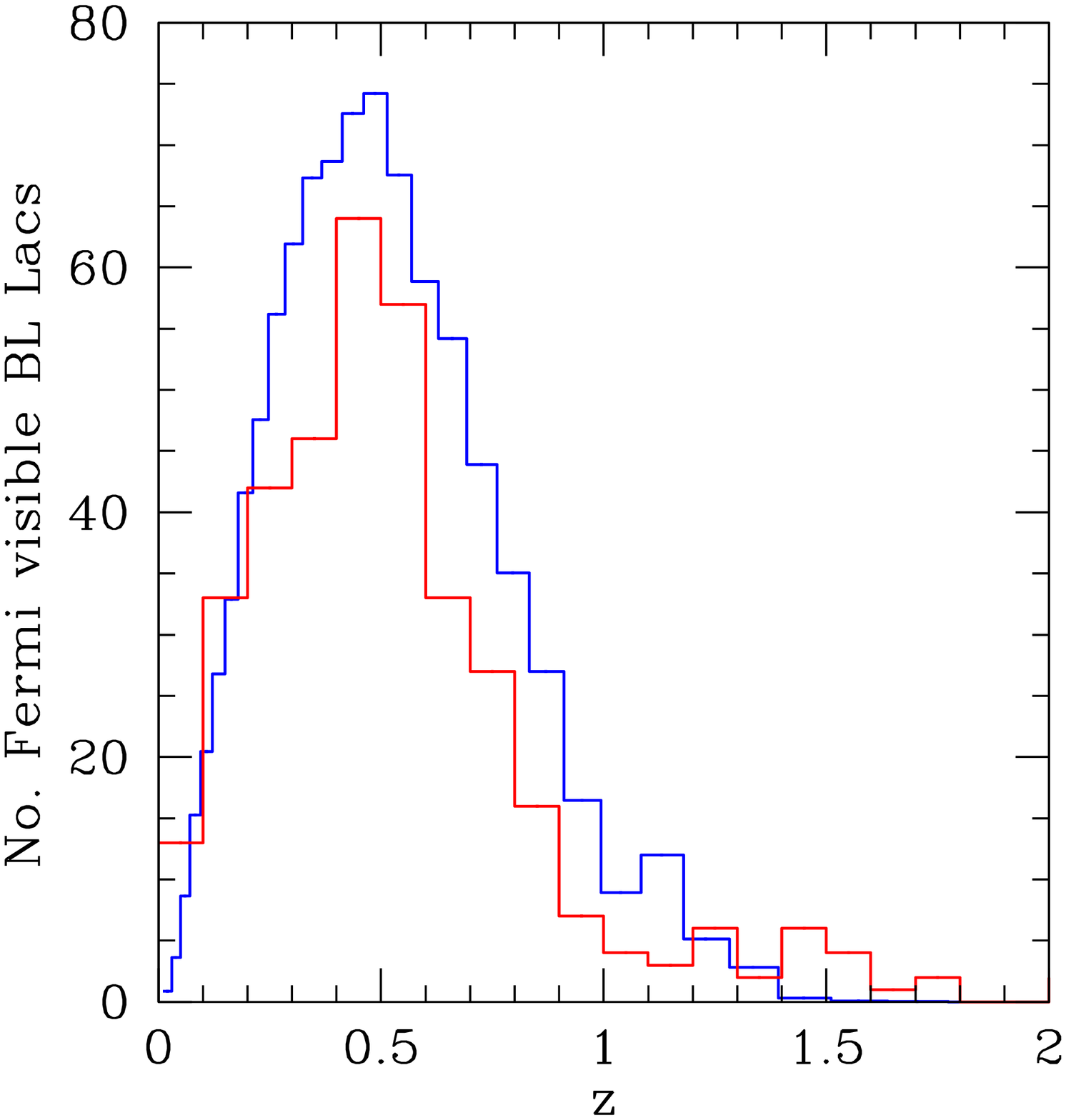} &
\epsfxsize=5cm \epsfbox{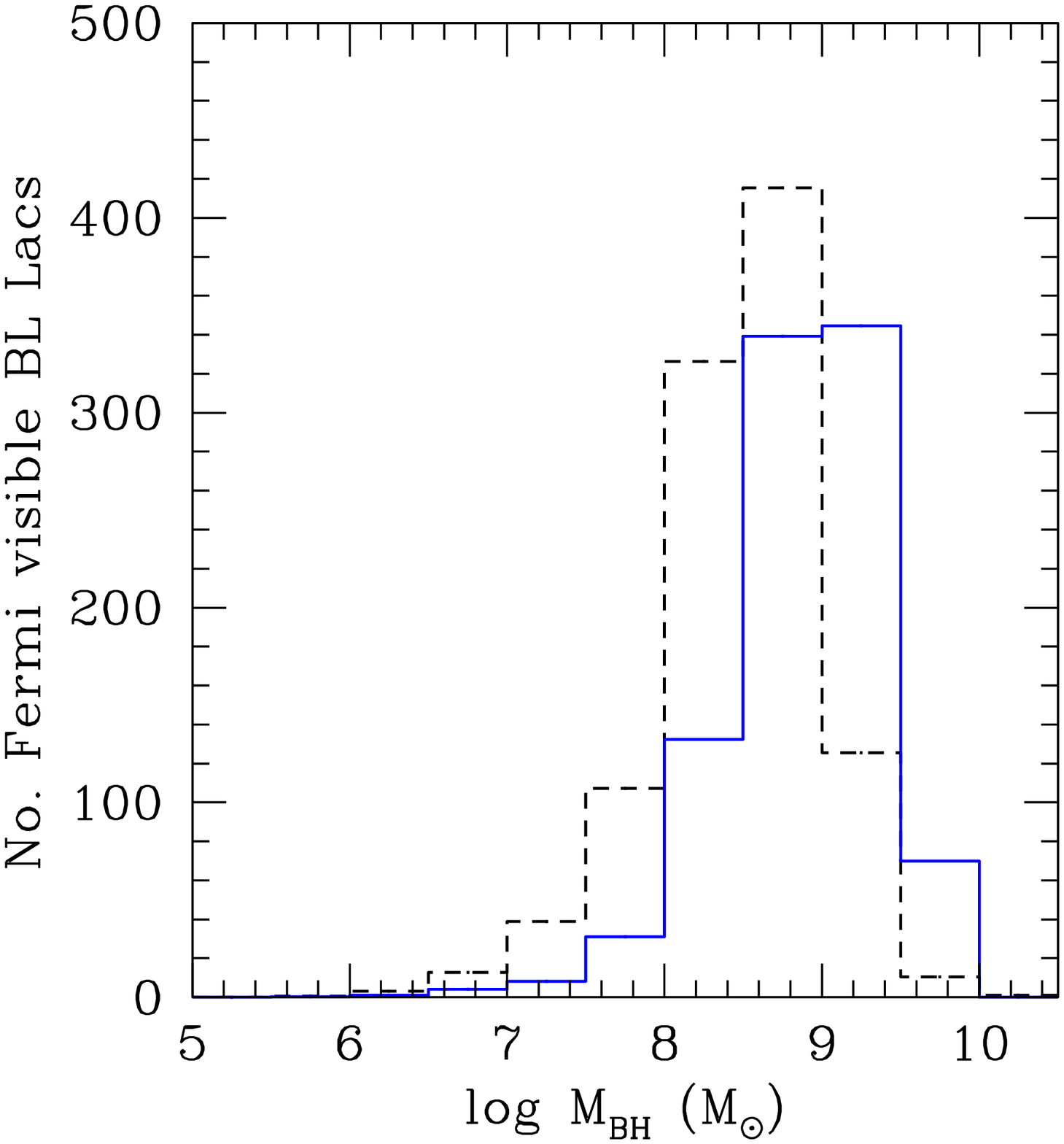} &
\epsfxsize=5cm \epsfbox{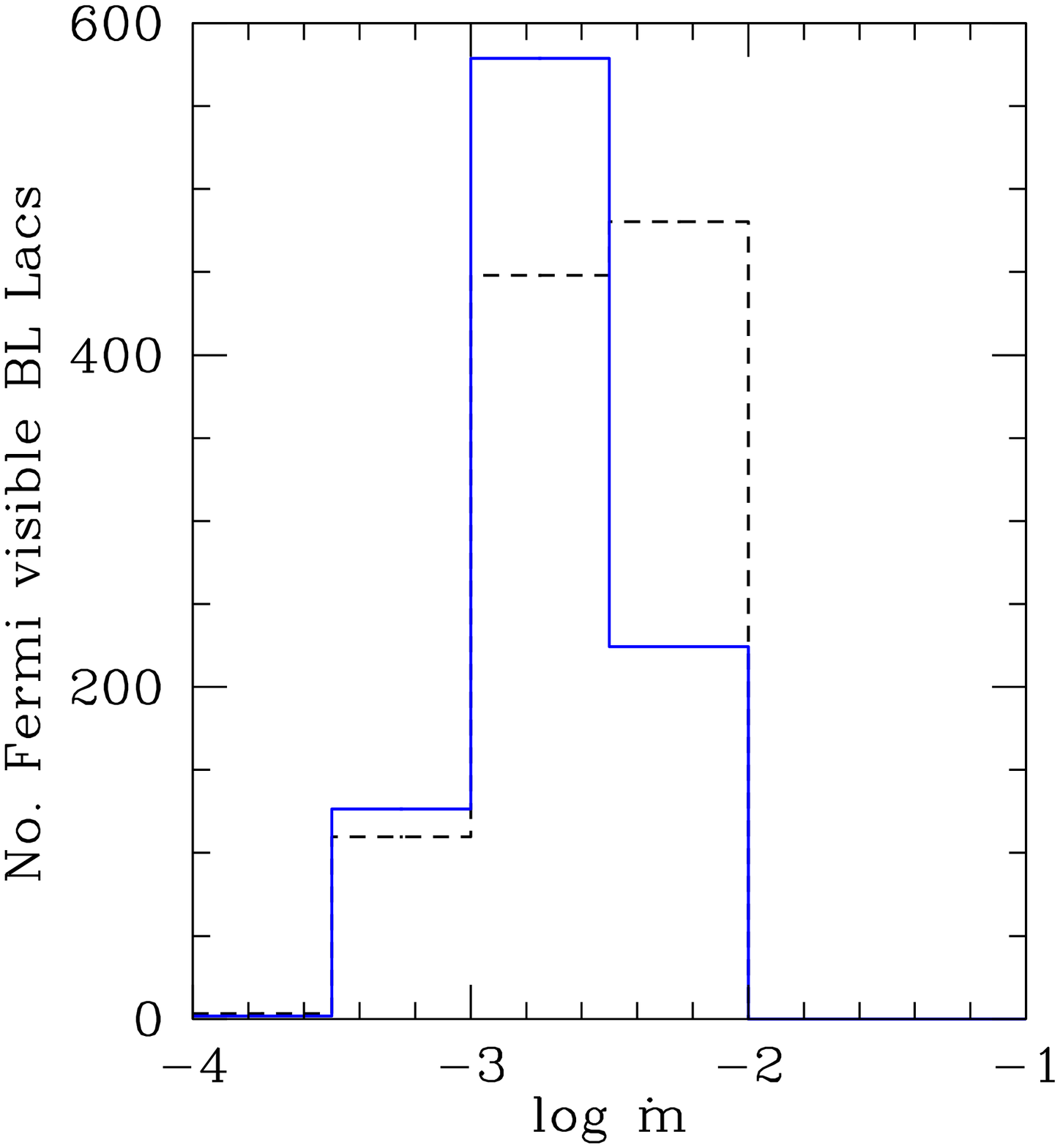} \\
\end{tabular}
\caption{a). Predicted redshift distribution of Fermi visible BL Lacs, assuming only BHs with spins $a>0.8$ produce a BL Lac type jet (blue). Red solid line shows observed redshift distribution of Fermi detected BL Lacs. b). Predicted mass distribution including only BHs with $a>0.8$ (blue). Black dashed line shows predicted distribution including all BHs $\times 0.01$. c). Predicted accretion rate distribution including only BHs with $a>0.8$ (blue). Black dashed line shows predicted distribution including all BHs $\times 0.01$.}
\label{fig1}
\end{figure*}

Cosmological simulations predict the number of supermassive black
holes accreting at different redshifts, together with their masses and
accretion rates. These simulations have been found to agree well with the
observed number densities of broad line and narrow line AGN in the
local universe (Fanidakis et al 2011; 2012). Combining our spectral
code with the black hole data from these simulations allows us to
predict the number of AGN that should be detected as BL Lacs by Fermi.

We combine our code with the black hole number densities predicted by
the Millennium Simulation (Springel et al 2005; Fanidakis et al 2011;
2012), binned as a function of both mass and mass accretion rate.  We
define a luminosity density from the number density multiplied by the
luminosity at that mass and mass accretion rate i.e. $L=\eta \dot{M}
c^2$ for the thin disc regime $10^{-2}<\dot{m}<1$, joining smoothly
onto a radiatively inefficient regime at lower $\dot{m}$ where
$L\propto \dot{m}^2$ (Narayan \& Yi 1995) and onto a super-Eddington
flow at higher $\dot{m}$ where $L\propto \ln(1+\dot{m})$ (Shakura \&
Sunyaev 1973). The luminosity density in each $(z,M,\dot{m})$ bin
therefore depends on the mass, accretion rate, spin (which sets
$\eta$), the inferred accretion regime and the number of black holes
in that bin.

Fig 4 shows the evolution of the luminosity density of accretion power
across cosmic time showing the features described by Fanidakis et al
(2011; 2012). At high redshift there is plenty of gas to fuel
accretion. The black holes accrete close to the Eddington limit and
grow rapidly. Comparing the snapshots for $z\sim9$ and $z\sim5$, the
typical black hole mass producing the bulk of the accretion luminosity
increases from $\sim10^6$ to $\sim10^8M_{\odot}$. As the black holes
gradually run out of gas, their accretion rates drop (compare $z\sim2$
and $z\sim1$). By redshift 2, accretion rates are beginning to drop
below $\dot{m}=10^{-2}$, into the regime at which BL Lac type jets
should be produced. This suggests no BL Lacs should be observed much
above $z\sim2$, not just because the flux becomes too faint, but
because the typical accretion rate is too high for the production of
BL Lac jets.

We select only black holes in the radiatively inefficient regime
($\dot{m}<10^{-2}$), assuming all BHs accreting inefficiently will
produce a BL Lac type jet, and calculate the number of AGN hosting a
BL Lac type jet in each $(z,M,\dot{m})$ bin.  If this number is less
than 1 we use Poisson statistics to randomly determine whether a BH is
present or not. Each black hole in each $(z,M,\dot{m})$ bin is then
assigned a random distance within this redshift bin and random
$\theta_{obs}$, assuming $\cos\theta_{obs}$ is distributed
uniformly. We then calculate the observed spectrum to determine
whether or not the jet would be visible to Fermi.

Fig 5a shows the predicted redshift distribution of Fermi visible BL
Lacs (black). The predicted distribution peaks at $z\sim 0.5$ and
drops gradually to $z\sim2$. No BL Lacs are observed above this point,
not because they are not visible (see Fig 3b), but because there
simply are not enough SMBHs accreting below $10^{-2}$ in the
cosmological simulations to produce SSC jets, due to the higher
activity expected at earlier times. The low redshift distribution of
BL Lacs is a direct result of cosmic downsizing and the requirement of
an $\dot{m}<10^{-2}$ to produce a SSC jet.

However, comparing this to observations shows a huge discrepancy (red
solid line from Shaw et al 2013, which almost merges with the X-axis
at this scale). Our expected number of $\sim 100000$ BL Lacs
dramatically overpredicts the observed number of Fermi detections
($\sim 500$). A clear illustration of the problem can be seen from
simply the number density of massive ($8<\log M <9$) black hole
accretion flows with $10^{-3}<\dot{M}<10^{-2}$ in the cosmological
simulations in the redshift bin centred around $z\sim 0.5$ (Fanidakis
et al 2011). This number is $6.8\times10^{-4} Mpc^{-3}$ and the volume
of this bin, from $0.509<z<0.564$ is $\sim9$ Gpc$^3$ so this gives
$6101659$ objects which should host similar jets to 1749+096 (Fig 1),
i.e. have Fermi flux of $10^{-10.5} (0.5/0.322)^{-2}\sim 10^{-11}$
ergs cm$^{-2}$ s$^{-1}$ if viewed at the same angle (roughly
$1/\Gamma$). The probability that we see the source within this angle
is $1-\cos(1/\Gamma)$, so the expected number of Fermi detections of
these sources alone is $\approx 13554$, similar to the full
calculation results.  The large discrepancy clearly points to a
fundamental breakdown of one of the assumptions.

On a more subtle level, the shape of the redshift distribution for the
Fermi predictions is also mismatched to the observations. The red
dashed line shows the observed number of BL Lacs scaled by a factor of
100 so it can be compared to the predicted distribution. We define
redshift from Shaw et al (2013) as either spectroscopic redshift,
spectroscopic lower limit, the mean of their redshifts derived from
host galaxy fitting, or their redshift upper limits, in that order of
preference. The dashed line shows this observed redshift
distribution ($\times 100$). It is clear that not only is the total
predicted number wrong, but we are also over estimating the proportion
of BL Lacs in the range $z=0.5-2$. 

The mass distribution and mass accretion rate distributions are as
expected (Figs 5b and c), with higher luminosity SSC flows
(i.e. higher $\dot{m} M$) being more likely to be observed, so simple
energetics selects the highest mass and mass accretion rate objects,
so the typical predicted mass of a Fermi visible BL Lac 
$\sim10^{8.5}-10^{9}M_{\odot}$ is a combination of three
factors:

\begin{enumerate}
\item Very few black holes accreting at $\dot{m}<10^{-2}$ above $z=2$.
\item Most black holes at $z<2$ accreting with $\dot{m}<10^{-2}$ have $M>10^8$.
\item Black holes with $M>10^9$ are increasingly rare in the local universe so we are less likely to observe one favourably orientated to our line of sight. 
\end{enumerate}

\section{Another Factor Affecting Jet Scaling?}

Our results predict $\sim10^5$ BL Lacs should have been detected in
the Fermi 2 year catalogue. In contrast, $\sim500$ objects in the 2nd
Fermi LAT catalogue are classed as BL Lacs. Even allowing for galactic
centre emission limiting sky coverage ($|b|>10^\circ$ means that only
80\% of the sky is included), this is still 3 orders of magnitude
larger than observed. Clearly jets do not simply scale with accretion
power.

We assumed the injected electron distribution was independent of mass
accretion rate. This may not be the case. Ghisellini \& Tavecchio
(2009) approximately use $\gamma_{max}\propto\gamma_b\propto1/\dot{m}$
to fit their blazar sequence. Acceleration of electrons is affected by
the ambient photon field which depends on the amount of cooling and
ultimately on accretion rate. In the efficient cooling regime,
$L_{sync}\propto L_{comp}\propto\dot{m}M$, making
$\gamma_{max}\propto1/\dot{m}$ not unreasonable. However, an increase
of $\gamma_{max}$, and $\gamma_b$ in particular, only serves to
increase the Fermi band luminosity for lower $\dot{m}$ systems, and
increase the discrepancy between the predictions and observations.

The discrepancy could instead be explained if every BH accreting below
$10^{-2}$ has the potential to produce a BL Lac type jet, but only
does so $1/1000th$ of the time. This seems unlikely, since Fanaroff
Riley Type I (FRI) AGN, the misaligned versions of BL Lacs (Padovani
\& Urry 1990, Padovani \& Urry 1991, Urry, Padovani \& Stickel 1991)
show large scale extended radio jets. This suggests these jets are
persistent, analogous to the steady low/hard state jet seen in BHBs at
low $\dot{m}$, not transient events. 

Another possibility is that the jet depends on magnetic
flux being advected onto the black hole from the extremely large scale
hot halo gas around the galaxy. Sikora \& Begelman (2013) suggest
that if there is magnetic flux in this gas, then it could be dragged
down close to the black hole by cold gas from a merging spiral
galaxy. However, this does not address the fundamental question as to
where the magnetic flux in the halo gas comes from, and 
using cold gas from a spiral merger to drag this field down to the
black hole is unlikely to be applicable in the BL Lacs as they have
low ongoing mass accretion rates. 

The bulk Lorentz factor of the jet is the biggest factor affecting its
visibility. We rerun our calculations with a reduced $\Gamma=10$
instead of 15 and find this roughly halves the predicted number of BL
Lacs, but still wildly overpredicts the observations.

We have assumed all jets are produced with the same value of bulk
Lorentz factor but this is clearly not the case - BHBs at low
$\dot{m}$ have $\Gamma\sim1.2$ (Fender et al 2004). The most obvious
way to reduce the number of visible BL Lacs is to allow a distribution
of $\Gamma$.  Yet there must be some physical parameter which controls
the jet acceleration. The acceleration region, where the magnetic
(Poynting) flux of the jet is converted to kinetic energy, is very
close to the black hole, so it seems most likely that this is set by
the black hole itself, in which case black hole spin is the only
remaining plausible parameter. A potential explanation for the lower
number of observed BL Lacs is that if only BHs with the highest spin
produce highly relativistic jets, and high spin is rare.

The cosmological simulations include the growth of SMBH spin via
accretion processes and black hole-black hole coalescence following
galaxy mergers (Volonteri et al 2005; 2007; 2012: Fanidakis et al 2011;
2012). The mass accumulated onto the central SMBH in an accretion
event is tied in the simulation to a fixed fraction (0.5\%) of the
mass of gas in a star formation episode in the host galaxy. If this
mass is all accreted in a single event (prolonged accretion) then this
is sufficient to spin most black holes up to maximal (Volonteri et al
2005; 2007). However, the mass accreting onto the central black hole
in any single event may be limited by self gravity. This splits the
accreting material up into multiple smaller events, each of which can
be randomly aligned since the star formation scale height is large
compared to the black hole sphere of influence even in a disc galaxy
(King et al 2008).  Such chaotic accretion flow models result in
predominantly low spin black holes (Volonteri et al 2007: King et al
2008: Fanidakis et al 2011; 2012), and high spins are rare as they are
produced not via accretion but via black hole mergers (Fanidakis et al
2011; 2012).

We use the spin distribution from the chaotic accretion flow model
simulations, and introduce a spin cut to our results, so that only BHs
with spin greater than $a_{cut}\sim 0.8$ produce a BL Lac type
jet. This reduces the predicted number of Fermi visible BL Lacs to
$\sim 900$. Fig 6a shows the resulting redshift distribution together
with the observed distribution. Not only does this reduce the
discrepancy between predicted and observed total numbers, it also
gives a better match to the shape of the distribution. Limiting
production of BL Lac type jets to BHs with high spin causes the
redshift distribution to peak slightly later and drop off more
sharply above $z=0.5$. This is because high spins arise from BH
mergers. Production of BL Lac jets is already limited to BHs accreting
below $10^{-2}$, ie. $M>10^8$. The BH mergers which make the most
massive BHs occurs at the latest times.

Fig 6b and 6c show how this affects the predicted mass and accretion
rate distributions of Fermi visible BL Lacs. The scaled down
distributions including BHs of all spins are shown by the dashed lines
for comparison. Requiring high spin increases the peak of the mass
distribution to $\sim 10^{8.5}-10^{9.5}$, because it is the most
massive BHs that are former by mergers and are consequently more
likely to have high spin. The peak of the accretion rate distribution
is actually slightly reduced. This is because the more massive BHs
have lower accretion rates; the spin cut has excluded lower mass BHs
with lower spins which tend to have slightly higher accretion rates.

The low spin, low accretion rate BHs, which generally have smaller
masses ($10^7-10^8 M_\odot$) correspond to the LINERs, which are not
observed to have jets as relativistic as those in BL Lacs. If they are
low spin, as expected, and high spin is required for a highly
relativistic jet, then this naturally explains why LINERs are observed
to have weaker radio jets.

\section{Implications of Scaling Jet Power with Spin: FRI Sources}

The parent population of BL Lacs is probably the Fanaroff Riley Type I
(FRI) sources (e.g. the review by Urry \& Padovani 1995). These show
'fluffy' radio jets whose surface brightness decreases with distance
from the central source, contrasting with the classic lobe and hotspot
radio emission seen in the more powerful FRII sources which are the
parent population of the FSRQ (Padovani \& Urry 1992). Thus the FRI
sources should also correspond to high spin BHs, and indeed are
similarly powered by high mass SMBHs (Woo \& Urry 2002).

However, we might then expect some difference in jet radio emission
between the FRIs and lower mass LINERs as the cosmological simulations
predict that the lower mass SMBHs have lower spin (Fanidakis et al
2011; 2012). Sikora et al (2007)
claim that this difference is indeed seen, with radio emission being
$\sim 3$ orders of magnitude higher in the FRIs. 

However, some of this difference disappears when only core radio
luminosity (rather than core plus lobes) is used (Broderick \& Fender
2011). This is clearly an issue as the extended radio emission must
depend on environment. The BHs in FRIs are more massive than those in
LINERs, hence live in richer cluster environments, with larger dark
matter halos which trap more hot cluster gas. The jet then emerges
into a denser, higher pressure environment, which means that a much
larger fraction (potentially all) of the jet kinetic energy is
converted to radiation and/or heating of the cluster gas (Birzan et al
2004). Conversely, any jet from the lower mass LINERs emerges into a
poorer group environment, so adiabatic losses can predominate and the
extended radio emission is much smaller (e.g. Krause et al 2012).

Some of the remaining difference in core radio power is {\em expected}
due to the difference in mass (Broderick \& Fender 2011). However, even
accounting for this there is still a factor of $\sim 10$ in mass
corrected, core radio emission. The LINERs lie on the Fundamental
plane (Merloni et al 2003; Falcke et al 2004) i.e. have the expected
core radio emission for their black hole mass and mass accretion rate, so
the FRIs are a factor 10 brighter in mass corrected, core radio
emission than expected from the same jet models which produce the low
bulk Lorentz factor BHB and LINER jets, consistent with the idea that
the jet is intrinsically more powerful/has higher Lorentz factor
due to black hole spin. 

It is difficult to predict the difference in core radio emission with
black hole spin in our models as synchrotron self-absorption means
that the observed radio emission does not arise in the same region as
produces the Fermi flux. It may be produced either at larger radii,
perhaps where the jet has decelerated, or in a lower density, lower
bulk Lorentz factor layer surrounding the $\Gamma=15$ spine of the
jet. Either of these could explain the lower Lorentz factor
($\Gamma\sim 2-10$) of the radio jet observed in FRIs (Chiaberge et al
2000), though the spine-layer structure may additionally be able to
explain the very fast variability timescales seen in some BL Lacs
(Ghisellini \& Tavecchio 2008). 

\section{Conclusions}

We have taken a statistical approach to constrain the conditions
necessary to produce the highly relativistic jets seen in BL Lac
objects. We combine SMBH number densities from cosmological
simulations, known to reproduce the optical luminosity function of
AGN, with spectral models of jet emission and simple jet scaling
functions which depend only on mass and accretion rate. The key
assumption is that every BH accreting with $\dot{m}<10^{-2}$ i.e. in
the radiatively inefficient accretion flow regime, should produce a BL
Lac type jet. 

Our calculation of the expected number of BL Lacs detectable by Fermi
overpredicts the observations by three orders of magnitude. This
clearly shows that our fundamental assumptions are incorrect, and that
the jet power and properties do not scale simply with mass and
mass accretion rate. The only other parameter which a black hole can
have is spin. We can reproduce the observed numbers of BL Lacs if SMBHs
grow predominantly via chaotic (randomly aligned) accretion episodes,
and that BL Lac type jets are restricted to BHs with spin $a>0.8$.
These are rare as they form from BH-BH coalescence following a major
merger event which is not then overwhelmed by further chaotic
accretion i.e. this requires a gas poor major merger event, and only
the most massive galaxies, which host the most massive black holes,
are gas poor in the local Universe (Fanidakis et al 2011; 2012).  

A spin cut is in line with the longstanding speculation that these
most relativistic jets require high spin black holes (Maraschi et al 2012), and also gives a
good match to the observed redshift distribution of BL Lacs which
peaks at $z=0.5$ and then drops off sharply, with no objects above
$z\sim2$. This is a consequence of three factors:

\begin{enumerate}
\item BL Lac jets are restricted to BHs with $\dot{m}<10^{-2}$, and there are no BHs accreting at $\dot{m}<10^{-2}$ above $z\sim2$.
\item Only the most massive BHs have high spin through mergers, which happen at late times, causing the bulk of the population to fall below $z=1$
\item These most massive objects are rare in the local universe causing the distribution to decrease again below $z=0.5$. 
\end{enumerate}

Since FRI sources are consistent with being the misaligned analogs of
BL Lacs, they should also have high spin. They are indeed offset from
the  Fundamental Plane, i.e. have higher (mass corrected) core radio
emission to the lower mass and presumably lower spin LINERs, though
only by a factor $\sim 10$ (Broderick \& Fender 2011). However, the
radio emission is not predominantly produced from the same region as
the Fermi flux so may not be as sensitive to the difference in spin. 

\section{Acknowledgements}

We thank Nikos Fanidakis and the Millennium simulation for use of their data. This work has made use of Ned Wright's Cosmology Calculator (Wright 2006). EG acknowledges funding from the UK STFC.

\appendix 
\section{}

The emission comes from a single zone of radius $R$. We assume material in the jet moves at a constant bulk Lorentz factor ($\Gamma$) and that some fraction of the transported electrons are accelerated into a power law distribution between minimum and maximum Lorentz factors $\gamma_{min}$ and $\gamma_{max}$, of the form: 

\begin{multline}
Q(\gamma)=Q_0\left(\frac{\gamma}{\gamma_b}\right)^{-n_1}\left(1+\gamma/\gamma_b\right)^{n1-n2}=Q_0q(\gamma) \\ 
\mbox{ for } \gamma_{min}<\gamma<\gamma_{max}
\end{multline}

$\gamma_b$ is the Lorentz factor at which the electron distribution changes in slope from $n_1$ to $n_2$. We calculate the normalisation $Q_0$ from the power injected into the accelerated electrons ($P_{rel}$):

\begin{equation}
P_{rel} = \frac{4\pi}{3} R^3m_ec^2Q_0\int_{\gamma_{min,inj}}^{\gamma_{max}}\gamma q(\gamma)d\gamma
\end{equation}

We calculate $\gamma_{cool}$ after a light crossing time $t_{cross}=R/c=\gamma_{cool}/\dot{\gamma}_{cool}$, as: 

\begin{equation}
\gamma_{cool}=\frac{3m_ec^2}{4\sigma_TRU_{seed}}
\end{equation}

Where $U_{seed}=U_B+U_{sync}$ is the sum of the energy density in magnetic fields and synchrotron emission which provides the seed photons for cooling. 

We solve the continuity equation to find the self consistent steady state electron distribution: 

\begin{equation}
\begin{split}
N(\gamma,t_{cross}) &= Kn(\gamma) \\
 &=
  \begin{cases} 
    AQ_0q(\gamma) \\ &\mbox{ for } \gamma_{min} < \gamma < \gamma_{cool} \\
    \frac{3m_ec^2}{4\sigma_TcU_{seed}}\frac{Q_0}{\gamma^2}\int_{\gamma}^{\gamma_{max}}q(\gamma)d\gamma \\ &\mbox{ for } \gamma_{cool} < \gamma <\gamma_{max} 
   \end{cases}
\end{split}
\end{equation}

Where $A$ is found by matching at $\gamma_{cool}$. 

We use the delta function approximation and calculate the synchrotron emissivity as: 

\begin{equation}
j_{sync}(\nu)=\frac{\sigma_Tc}{6\pi\nu_B}U_B\gamma N(\gamma)
\end{equation}

Where the electron Lorentz factor and synchrotron photon frequency are related by $\gamma=\sqrt{3\nu/4\nu_B}$ and we calculate the synchrotron self-absorption frequency ($\nu_{ssa}$) as given by (Ghisellini et al. 1985):

\begin{equation}
\nu_{ssa}=\left(4.62\times10^{14}KB^{2.5}\frac{R_j}{0.7}\right)^{2/7}
\end{equation}

We calculate synchrotron self-Compton emission including the Klein-Nishina cross section using the delta approximation:

\begin{equation}
j_{comp}(\nu)=\frac{\sigma_Tc}{6\pi}\int^{\gamma_{max}}_{\gamma_{min}}\int^{\nu_{sync,max}}_{\nu_{ssa}}\frac{U_{sync}(\nu_{sync})}{\nu_{sync}}\gamma N(\gamma)d\nu_{sync}d\gamma
\end{equation}

Where electron Lorentz factor and Compton photon frequency are related by $\gamma=\sqrt{3\nu/4\nu_{sync}}$. 

Bulk motion of the jet boosts and blue shifts the emission. We calculate the observed flux as: 

\begin{equation}
F(\nu\delta/(1+z))=\frac{(j_{sync}(\nu)+j_{comp}(\nu))}{R_{co}^2}\frac{4\pi}{3} R^3\delta^3
\end{equation}

Where $\delta=(\Gamma-\cos{\theta}\sqrt{\Gamma^2-1})^{-1}$ is the Doppler factor and $R_{co}$ is the comoving distance to the object at redshift $z$. 

\label{lastpage}

\end{document}